\documentclass[twocolumn,showpacs,preprintnumbers,amsmath,amssymb,prb]{revtex4}
\usepackage{graphicx}
\usepackage{dcolumn}
\usepackage{bm}

\begin{document}

\title{
Scaling Analysis in the Numerical
Renormalization Group Study of the Sub-Ohmic Spin-Boson Model }

\author{Ning-Hua Tong}
\email{nhtong@ruc.edu.cn}
\affiliation{Department of Physics, Renmin University of China, 100872 Beijing, China}

\author{Yan-Hua Hou}
\email{phy.hyh@gmail.com}
\affiliation{Department of Physics, Renmin University of China, 100872 Beijing, China}
\date{\today}

\begin{abstract}
The spin-boson model has nontrivial quantum phase transitions in the
sub-Ohmic regime. For the bath spectra exponent $0 \leqslant s<1/2$,
the bosonic numerical renormalization group (BNRG) study of the
exponents $\beta$ and $\delta$ are hampered by the boson-state
truncation, which leads to artificial interacting exponents instead
of the correct Gaussian ones. In this paper, guided by a mean-field
calculation, we study the order-parameter function
$m(\tau=\alpha-\alpha_c, \epsilon, \Delta)$ using BNRG. Scaling
analysis with respect to the boson-state truncation $N_{b}$, the
logarithmic discretization parameter $\Lambda$, and the tunneling
strength $\Delta$ are carried out. Truncation-induced multiple-power
behaviors are observed close to the critical point,
 with artificial values of $\beta$ and $\delta$. They cross over to
classical behaviors with exponents $\beta=1/2$ and $\delta=3$ on the
intermediate scales of $\tau$ and $\epsilon$, respectively. We also
find $\tau/\Delta^{1-s}$ and $\epsilon/\Delta$ scalings in the
function $m(\tau, \epsilon, \Delta)$. The role of boson-state
truncation as a scaling variable in the BNRG result for $0 \leqslant
s<1/2$ is identified and its interplay with the
 logarithmic discretization revealed. Relevance to the validity of
 quantum-to-classical mapping in other impurity models is discussed.
\end{abstract}

\maketitle
\section{Introduction}

The spin-boson model is the simplest model that describes a quantum
two-level system subjected to the influence of a dissipative
environment. It has applications in various fields in
physics~\cite{Weiss1}
 and its properties are studied extensively.~\cite{Leggett1}
Especially, for the bath spectra exponent $0 \leqslant s<1$
(sub-Ohmic bath), the ground state of the spin-boson model may
change from the spin-tunneling state to the spin-pinned state
through a second-order phase transition, as the dissipation strength
crosses a critical value from below. This environment-induced
quantum phase transition attracts much attention in the past few
years.~\cite{Kehrein1, Bulla1, Vojta1, Tong1, Bulla2}

Among the many theoretical methods that have been used to study this
quantum phase transition, the bosonic numerical renormalization
group (NRG) method is regarded as the most accurate one.
Technically, NRG is composed of three standard procedures:
logarithmic discretization, transforming the Hamiltonian into a
semi-infinite chain, and the iterative diagonalization. Thanks to
the logarithmic discretization, the state information at
exponentially small energy scales is kept along the iterative
diagonalization. Therefore, NRG allows for reliable extraction of
critical exponents and description of the crossover behavior. Two
parameters control the precision of NRG, i.e., the logarithmic
discretization $\Lambda$ ($\Lambda>1$) and the number of kept states
$M_s$. The original Wilson's NRG is designed for the impurity
problem with a fermionic bath.~\cite{Wilson1} In the past few years,
the extension of NRG to impurity models with bosonic bath, such as
the spin-boson model, proved to be
fruitful.~\cite{Bulla1,Lee1,Tornow1} Due to the infinite local
Hilbert space for each bosonic bath mode, one has to truncate the
space into $N_b$ states. Usually the boson occupation states are
used as the local bases, although optimal bases have also been
considered.~\cite{Bulla1} Using this bosonic NRG (BNRG), the
thermodynamical as well as dynamical quantities for the spin-boson
model are studied and the critical exponents in the sub-Ohmic regime
obtained. Although it was clear that the localized fixed point
($\langle \sigma_z\rangle \not= 0$) cannot be described exactly due
to the truncation of boson states, it was believed and partly
checked~\cite{Bulla1,Vojta1} that the BNRG shares the virtue of NRG,
i.e, the parameters $\Lambda$, $M_s$, and $N_b$ only influences the
value of non-universal quantities, such as the critical point value
$\alpha_c$ and the prefactors of power laws. The exact value of them
can be reliably obtained by extrapolating $\Lambda \rightarrow 1$,
$M_s \rightarrow \infty$, and $N_b \rightarrow \infty$. The
universal quantities such as the critical exponents are not supposed
to depend on $\Lambda$, $M_{s}$, or $N_{b}$.

For the quantum critical behavior of the spin-boson model in the
sub-Ohmic regime, the following critical exponents have been
studied~\cite{Bulla1,Vojta1}:

\begin{eqnarray}
  && m (\alpha > \alpha_c, T=0, \epsilon=0) \propto \left(\alpha - \alpha_c
  \right)^{\beta} \!\! \nonumber\\
  && \nonumber\\
  &&  m (\alpha = \alpha_c, T=0, \epsilon) \propto
   \epsilon^{1/\delta} \!\! \nonumber\\
   && \nonumber\\
  &&\chi (\alpha, T=0, \epsilon=0) \propto |\alpha-\alpha_c|^{-{\gamma}} \nonumber \\
  && \nonumber\\
  &&  T^{*}(\alpha, \epsilon=0) \propto |\alpha-\alpha_c|^{z \nu}   \nonumber \\
  && \nonumber\\
  &&  \chi(\alpha=\alpha_c, T, \epsilon=0) \propto T^{-x} \nonumber \\
  && \nonumber\\
  &&  C(\omega)(\alpha=\alpha_c, T=0, \epsilon=0) \propto \omega^{-y}
\end{eqnarray}

Here, $\alpha$ and $\epsilon$ describe the coupling strength between
the spin and bosons, and the bias field on the spin, respectively.
$\chi$ and $C(\omega)$ are static susceptibility and dynamical spin
correlation function, respectively. The naive BNRG study of
these critical exponents leads to the following conclusions.~\cite{Bulla1,Vojta1}
(i) In the regime $0 \leqslant s< 1$, the critical fixed points are interacting and
the corresponding critical exponents are non-classical; and (ii) The
hyperscaling relation and $\omega/T$ scaling hold. In the regime
$0 \leqslant s<1/2$, these conclusions are in contrast to previous theories
based on the quantum-to-classical mapping. There, the spin-boson
model is mapped into a one-dimensional Ising model with $J_{ij}
\propto 1/(r_i-r_j)^{(1+s)}$.~\cite{Leggett1} In the regime
$0\leqslant s<1/2$, this Ising model is above its upper critical dimension,
leading to Gaussian critical fixed point and classical exponents.
The hyperscaling relation and $\omega/T$ scaling do not hold
there.~\cite{Dyson1,Fisher1,Luijten1} Recently, using
a number of new methods,~\cite{Winter1,Alvermann1,Wong1,Zhang1,Lv1}
the quantum phase transition in the sub-Ohmic spin-boson model has been studied.
The obtained critical coupling strength
$\alpha_c(s)$,\cite{Winter1,Alvermann1,Wong1,Zhang1,Lv1} the
exponent $\nu$,~\cite{Winter1} and $\gamma$ (Ref.~\onlinecite{Alvermann1}) are
consistent with the BNRG results. However, the quantum Monte Carlo
(QMC) simulation~\cite{Winter1} and the exact diagonalization
study~\cite{Alvermann1} found that in the regime $0 \leqslant s<1/2$, the
critical point is Gaussian with classical exponent $\beta=1/2$,
being different from the BNRG conclusion. In the regime $1/2 \leqslant s<
1$, BNRG results are consistent with the quantum-to-classical
mapping theory which predicts an interacting fixed point and
$\omega/T$ scaling.

Recently, a closer examination of BNRG method discloses two sources
of error, which were not noticed before. One is the boson state
truncation error,~\cite{Vojta2} the other is the mass-flow
error.~\cite{Vojta2,Vojta3} The boson state truncation error spoils
the evaluation of the order-parameter related exponents $\beta$ and
$\delta$, while the mass-flow problem influences the correct
evaluation of $x$. The two sources of error are different in nature
and exist simultaneously in the BNRG algorithm in the whole regime
$0\leqslant s<1$. But, they influence the critical behavior only in
the regime $0\leqslant s<1/2$, where the critical fixed point is
expected to be Gaussian in the absence of these errors.

For the mean-field spin-boson model which has a Gaussian critical
fixed point,~\cite{Hou1} we showed that the boson state truncation
leads to an artificial interacting fixed point in the regime $0
\leqslant s < 1/2$, but has no influence in $1/2 \leqslant s < 1$.
This is an example where the boson-state truncation destroys the
Gaussian fixed point and spoils the correct calculation of the
exponents $\beta$ and $\delta$. It leads to the surmise that the
same may happen in the BNRG study of the spin-boson model. It would
be difficult to find the Gaussian nature of the critical fixed point
in $0\leqslant s<1/2$ using BNRG, if the truncation works the same
way as in the mean-field Hamiltonian.

In Ref.~\onlinecite{Vojta2}, the boson-state truncation error is
traced back to the presence of a dangerously irrelevant variable for
a Gaussian critical fixed point. The correct exponent can be seen on
the intermediate scales. For the more fundamental problem of mass-
flow error, Vojta {\it et al.}~\cite{Vojta3} have proposed an
extended NRG algorithm to partly remedy the problem and got the
correct exponent $x=1/2$ in $0\leqslant s<1/2$. In this paper, we
focus on the boson state truncation error. In the BNRG, it is still
unclear how a finite $N_{b}$ leads to wrong $\beta$ and $\delta$,
and how to extract the correct exponents. For the spin-boson model,
a thorough numerical study in the regime $0 \leqslant s < 1/2$ is
required to prove or disprove the validity of the
quantum-to-classical mapping in this
model.\cite{Winter1,Alvermann1,Vojta2,Vojta3,Kirchner1,Florens1}
Here, we use the scaling approach to analyze the BNRG data with
respect to boson-state truncation $N_b$, logarithmic discretization
parameter $\Lambda$, and tunneling strength $\Delta$. We find that
for any finite $N_{b}$,
 the order parameter $m$ has a multiple power form like that at the tricritical
point,~\cite{Hankey1} with nonclassical exponents $\beta$ and
$\delta$ dominated by the discretization scheme. The correct power-
law behavior can be observed on the intermediate scale away from the
critical point. These two different power-law regimes are connected
at a crossover scale, which goes to zero as a power of $x=1/N_{b}$
and $w=\Lambda-1$. Thus, in the limit of either $x \rightarrow 0$ or
$w \rightarrow 0$, the classical critical exponents $\beta$ and
$\delta$ are recovered. This is the same as in the mean-field
spin-boson model, which we will detail in the Appendix. Besides, we
also disclose the role of $\Delta$ as a scaling variable in the
order
 parameter close to criticality.

In Sec. II, the spin-boson model and our main results are presented.
In Sec. III, a summary and discussion will be made. In the Appendix,
we present the critical behavior of order parameter and
susceptibility for the mean-field spin-boson model.

\section{Model and Results}

The Hamiltonian of the spin-boson model reads as
\begin{equation}
  H=-\frac{\Delta}{2} \sigma_x + \frac{\epsilon}{2} \sigma_z
  +\frac{1}{2}\sigma_z  \sum_{i} \lambda_{i}\left(a_{i}+a_{i}^{\dagger} \right)
  + \sum_{i} \omega_{i} a_{i}^{\dagger}a_{i}.
\end{equation}
Here, $\sigma_x$ and $\sigma_z$ are Pauli matrices, and $a_{i}$ and
$a_{i}^{\dagger}$ are the bosonic annihilation and creation
operators of the mode $i$, respectively. The properties of the
quantum two level system are determined by the environment through
the following bath spectrum~\cite{Leggett1}:
\begin{equation}
   J(\omega) =\pi \sum_{i} \lambda_{i}^{2} \delta(\omega -
   \omega_{i}),
\end{equation}
for which we assume the following power form,
\begin{equation}
   J(\omega)= 2\pi\alpha \omega^{s} \omega_{c}^{1-s}, \,\,\,\,\, (0< \omega <
   \omega_c).
\end{equation}
Here, $\alpha$ measures the strength of the dissipation.
$\omega_c=1$ is used as the energy unit.

\subsection{Mean-field results}
Before we carry out NRG calculations, it would be helpful to first
have a look at the mean-field spin-boson model. It has Gaussian
critical fixed point and classical exponents for any $s \geqslant
0$. It is used to mimic the situation of the full spin-boson model
in the regime $0\leqslant s<1/2$.
 In Ref.~\onlinecite{Hou1}, the scaling behavior of $m={\langle
\sigma_z\rangle}$ with respect to boson state truncation $N_{b}$ was
investigated numerically. It was found that any finite $N_{b}$ will
lead to non-mean-field exponents $\beta$ and $\delta$ in the regime
$0<s<1/2$ and an exponential behavior at $s=0$. The exponents (as
functions of $s$) agree well with those extracted from BNRG
calculations for the full spin-boson model. Here, we present a
concise and complete analytical solution which will guide our BNRG
study in the next section.

The Hamiltonian of the mean-field spin-boson model reads as
(neglecting a constant),
\begin{eqnarray}
  H=&&-\frac{\Delta}{2} \sigma_x + \frac{\epsilon}{2} \sigma_z +\frac{1}{2} \sigma_z \sum_{i} \lambda_{i}\langle a_{i}+a_{i}^{\dagger} \rangle
  \nonumber \\
   && +\frac{1}{2}\langle \sigma_z \rangle  \sum_{i} \lambda_{i}\left(a_{i}+a_{i}^{\dagger} \right)
    + \sum_{i} \omega_{i} a_{i}^{\dagger}a_{i}.
\end{eqnarray}
To make connection with the NRG study, we carry out a standard
logarithmic discretization~\cite{Wilson1,Bulla2}. The obtained
star-type mean-field Hamiltonian reads as
\begin{eqnarray}\label{eq:8}
  H_{mf}^{star}= &&  -\frac{\Delta}{2}\sigma_{x}+ \left[ \frac{\epsilon}{2}+ \frac{1}{2\sqrt{\pi}} \sum_{n}\gamma_{n}
   \langle a_{n}+a_{n}^{\dag} \rangle \right] \sigma_{z} \nonumber   \\
    && + \sum_{n}\xi_{n}a_{n}^{\dag}a_{n} + \frac{\langle \sigma_z\rangle}{2\sqrt{\pi}}
   \sum_{n}\gamma_{n}\left(a_{n}+a_{n}^{\dag} \right).
\end{eqnarray}
Here, the logarithmic discretization gives
\begin{eqnarray}\label{eq:11}
\gamma_{n}^{2}=\frac{2\pi\alpha}{1+s}\left[1-\Lambda^{-(1+s)}\right]
\Lambda^{-n(1+s)}\omega_{c}^{2},
\end{eqnarray}
and
\begin{eqnarray}\label{eq:12}
\xi_{n}=\frac{1+s}{2+s}\frac{1-\Lambda^{-(2+s)}}{1-\Lambda^{-(1+s)}}\Lambda^{-n}\omega_{c}.
\end{eqnarray}
$\Lambda >1$ is the logarithmic discretization parameter.

For the spin and boson decoupled Hamiltonian Eq.(6), the
self-consistent equations are easily solved when there is no
truncation, i.e., $N_{b}=\infty$. For finite $N_{b}$, this model
cannot be solved exactly. However, through an analysis of the
single-mode Hamiltonian, we get the asymptotically exact expression
for the order parameter $m$ as a function of $\alpha - \alpha_{c}$
and $\epsilon$ for given $N_{b}$, $\Lambda$, and $\Delta$. We
summarize the results in the following and leave the detailed
derivation in the Appendix.

For a fixed $\Delta$, the critical $\alpha_c$ for Eq.(6) does not
depend on $N_{b}$, but only on $\Lambda$. We define the following
scaling variables $\tau=\alpha-\alpha_c(\Lambda)$, $x=1/N_{b}$, and
$w=\Lambda-1$. To obtain the exponent $\beta$, we take $\epsilon=0$.
Magnetization $m(\tau, \epsilon=0, x, w)=\langle
\sigma_{z}\rangle/2$ has the following behavior in the limit $\tau,
x, w \rightarrow 0$. For $s \geqslant 1/2$,
\begin{equation}
   m(\tau, \epsilon=0, x, w)=1/\sqrt{2\alpha_c(1)}\tau^{1/2}.
\end{equation}
For $0<s<1/2$, we get
\begin{eqnarray}
&&m(\tau, \epsilon=0, x, w)   \nonumber \\
&& = \left\{
\begin{array}{lll}
(\frac{2}{c})^{\frac{1-s}{2s}}[\alpha_c(1)]^{\frac{s-2}{2s}}\left(\frac{\Delta}{\omega_c}
\right)^{\frac{1-s}{2s}}(w x)^{-\frac{1}{2}}\tau^{\frac{1-s}{2s}} ,\,\,   & (\tau \ll \tau_{cr}); \\
& \\
 1/\sqrt{2 \alpha_c(1)} \tau^{\frac{1}{2}} ,\,\, & (\tau \gg \tau_{cr}).
\end{array} \right. \nonumber \\
&&
\end{eqnarray}
Here, $\alpha_{c}(1)=\Delta s /(2 \omega_c)$ is the mean-field
critical point for $\Lambda=1$. The crossover scale $\tau_{cr}$
reads as
\begin{equation}
\tau_{cr}=c^{\prime}[\alpha_{c}(1)]^{\frac{2(1-s)}{1-2s}}\left(
\frac{\Delta}{\omega_c} \right)^{ \frac{1-s}{2s-1}} \left(
wx\right)^{ \frac{s}{1-2s} }.
\end{equation}
For $s=0$, we get $\alpha_c(1)=0$ and the exponential behavior
\begin{equation}
  m(\tau, \epsilon=0, x ,w)= (4w x \tau)^{-1/2} e^{-\frac{\Delta}{4\omega_c
  \tau}}.
\end{equation}
If one takes $x=0$ before $s \rightarrow 0$ is taken, one gets a
singular behavior, $m(\tau, \epsilon=0, x=0 ,w)=1/2$ for $\tau
>0$ and $m(\tau, \epsilon=0, x=0 ,w)=0$ for $\tau=0$.

To obtain the exponent $\delta$, we take $\tau=0$. In the limit
$\epsilon, x, w \rightarrow 0$, we get for $s\geqslant 1/2$,
\begin{equation}
   m(\tau=0, \epsilon, x, w)=- \left( \frac{\epsilon}{4 \Delta} \right)^{1/3}.
\end{equation}
For $0<s<1/2$, we get
\begin{eqnarray}
&&m(\tau=0, \epsilon, x, w)   \nonumber \\
&& = \left\{
\begin{array}{lll}
- c^{-\frac{1-s}{1+s}} [\alpha_c(1)]^{-\frac{1}{1+s}} (w
x)^{-\frac{s}{1+s}}(\frac{\epsilon}{\omega_c})^{\frac{1-s}{1+s}}
  ,\,\,\,\,\,\,\,\,\,   & (\epsilon \ll \epsilon_{cr} ); \\
&\\
 - (\frac{\epsilon}{4 \Delta})^{1/3} ,\,\,\,\,\,\,\,\,\, &(\epsilon \gg \epsilon_{cr}).
\end{array} \right. \nonumber \\
&&
\end{eqnarray}
The crossover bias $\epsilon_{cr}$ in Eq.(14) reads as
\begin{equation}
\frac{\epsilon_{cr}}{\omega_c}=
c^{\prime\prime}[\alpha_{c}(1)]^{\frac{3}{2(1-2s)}}\left(
\frac{\Delta}{\omega_c} \right)^{ \frac{1+s}{2(2s-1)}} \left(
wx\right)^{ \frac{3s}{2(1-2s)} },
\end{equation}
At $s=0$, we have
\begin{equation}
m(\tau=0, \epsilon, x, w) =-\frac{\epsilon}{2 \Delta}.
\end{equation}
In the equations above, $c$, $c^{\prime}$, and $c^{\prime\prime}$
are constants which are independent of $x$, $w$, and $\Delta$.

These expressions are consistent with the numerical solution of the
mean-field spin-boson model and subsequent $N_{b}$-scaling analysis
(the exponent for $x$ fitted in Ref.~\onlinecite{Hou1} deviates due
to numerical errors.). A new result here is that $w$ also becomes a
scaling variable. This implies that the logarithmic
 discretization can no longer keep the universal properties intact in the regime $0
\leqslant s < 1/2$ as is usually assumed in the NRG studies.

The above results clearly show that in the regime $0 \leqslant s <
1/2$, boson state truncation indeed overtakes the critical fixed
point of the mean-field spin-boson model, changing it from Gaussian
to interacting. For any finite $N_{b}$, we get the exponents
$\beta=(1-s)/(2s)$ and $\delta=(1+s)/(1-s)$ as long as $\tau$ or
$\epsilon$ is sufficiently small. A remarkable observation is that
these expressions agree well with $\beta_{NRG}$ and $\delta_{NRG}$,
 the BNRG exponents for the full spin-boson model with finite $N_b$
 (See Fig.14(a) and (b)).~\cite{Hou1} Hence we have the following relations,
\begin{eqnarray}
&& \beta_{NRG}=(1-s)/(2s)  \,\,\,\, , \nonumber \\
&& \delta_{NRG}=(1+s)/(1-s)  \,\,\,\,.
\end{eqnarray}
In contrast, the exponents $\beta_{MF}=1/2$ and
$\delta_{MF}=3$ only appear when $\tau$ and $\epsilon$ are larger
than their respective crossover scales.

It is noted that for a finite $N_{b}$, the result of the mean-field
Hamiltonian Eq.(5) depends on the parametrization used for the bath
degrees of freedom. Expressions (9)-(16) hold only for
$H^{star}_{mf}$ in Eq.(6), which is obtained from a specific
parametrization, i.e., the logarithmic discretization. For other
parametrization schemes, different exponents will obtain. At
$N_{b}=\infty$, bosons become canonical and one gets the exact
classical exponents irrespective of the parametrization of the bath.

The above discussions are for the mean-field star-Hamiltonian
obtained from logarithmic discretization. The mean-field
chain-Hamiltonian cannot be solved analytically at finite $N_{b}$.
Using BNRG, we managed to solve the mean field equations by
iteration. We found that the converged solution has no qualitative
difference from that of the mean-field star-Hamiltonian. That is,
the same nonclassical (classical) exponents are obtained in the low
(high) energy regime.

\begin{figure}[t!]
\begin{center}
\includegraphics[width=2.7in, height=3.9in, angle=270]{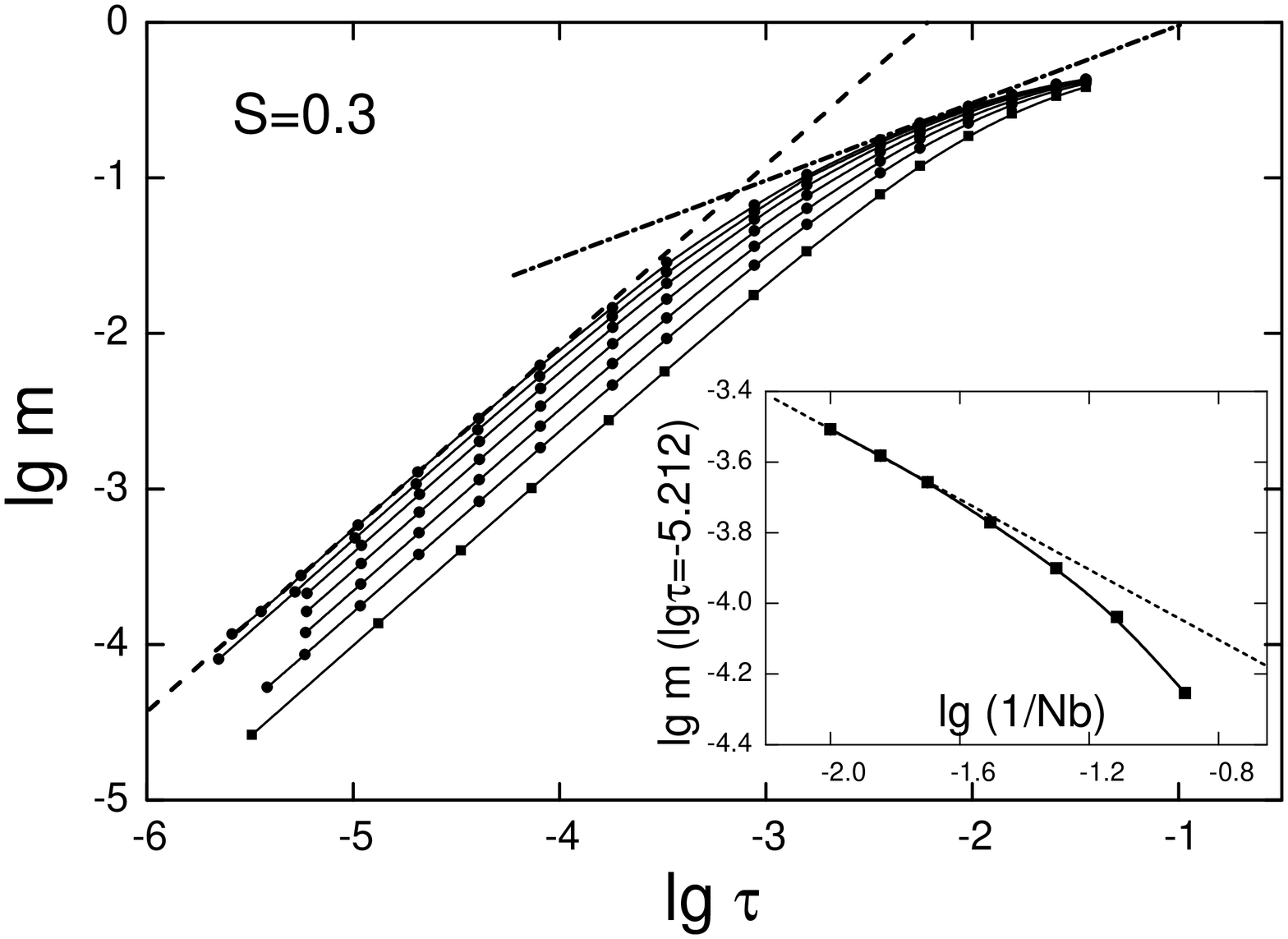}
\vspace*{-1.2cm}
\end{center}
\caption{$\text{lg}m(\epsilon=0)$ vs. $\text{lg}\tau$ at $s=0.3$ for
different $N_{b}$ values. Throughout this paper, $\text{lg}$ stands
for logarithm base 10.
 From bottom to top, $N_{b}=8, 13, 20, 32, 50, 70, 100$, respectively. Dots with
solid guiding lines are BNRG data. The dashed line is the power law
fit for $N_{b}=100$ in small $\tau$ regime, which gives
$\beta_{NRG}=1.169$ being consistent with $(1-s)/(2s)=1.167$. The
guiding dash-dotted line marks the $\beta=1/2$ behavior in the large
$N_{b}$ limit. Other parameters are $\Delta=0.1$, $\Lambda=10.0$,
and $M_{s}=80$. Inset: $\text{lg}m(\epsilon=0)$ as a function of
$1/N_{b}$ at $\text{lg}\tau=-5.212$.
 The guiding dashed line with slope $-1/2$ shows the asymptotic
power law in the large $N_{b}$ limit.}
\end{figure}
\subsection{NRG results for $s=0.3$}

In this section, we present the BNRG data for $m(\tau, \epsilon,
\Delta, x, w)$, carry out scaling analysis, and extract the
exponents $\beta$ and $\delta$ in the limit $N_{b}=\infty$. The
scaling analysis is in parallel with that in Ref.~\onlinecite{Hou1}.
We will demonstrate that in the regime $0<s<1/2$, BNRG data fulfills
Eqs.(10) and (14), except that $\alpha_c(1)$ there should be
replaced with the corresponding BNRG values.

For simplicity, we focus on $s=0.3$, a generic value in the regime
$0 < s < 1/2$. Extensive BNRG calculations are done with various
parameters $N_{b}$, $\Lambda$, and $M_{s}$. Due to computational
limitations, we use $N_{b}=8 \sim 100$, $\Lambda=2 \sim 10$, and
$M_{s}=60 \sim 120$. We define $\tau=\alpha-\alpha_c(\Delta,
\Lambda, N_{b}, M_s)$. $\alpha_c(\Delta, \Lambda, N_{b}, M_{s})$ is
the critical dissipation strength for a fixed set of parameters. We
found that $\alpha_c(\Delta, \Lambda, N_{b}, M_s)$ has almost no
dependence on $N_{b}$ (less than $10^{-4}$ percent between
$N_{b}=13$ and $N_{b}=50$), similar to the mean-field case where
$\alpha_c$ strictly does not depend on
 $N_{b}$. For $\Lambda> 2$ and $N_{b}<100$, $\alpha_c$ converges very fast as $M_{s}$
increases. For $\Delta$ and $\Lambda$ dependence, we find
\begin{eqnarray}
\alpha_c(\Delta, \Lambda, N_{b}, M_s) \propto \left\{
\begin{array}{lll}
 \Delta^{1-s} ,\,\,\,\,\,   & (\Delta \rightarrow 0); \\
&\\
 \alpha_c(1)+ c (\Lambda-1) ,\,\,\,\,\, & (\Lambda \rightarrow 1).
\end{array} \right.
\end{eqnarray}
Here $c \approx 0.003$ for $N_{b}=8$, $M_{s}=80$ is quite small.
This explains the very good agreement between the BNRG curve
$\alpha_{c}(s)$ and that from other
methods.~\cite{Winter1,Alvermann1,Wong1,Zhang1,Lv1}

\subsubsection{$m(\tau, \epsilon=0, \Delta, x, w)$}

\begin{figure}[t!]
\begin{center}
\includegraphics[width=2.7in, height=3.9in, angle=270]{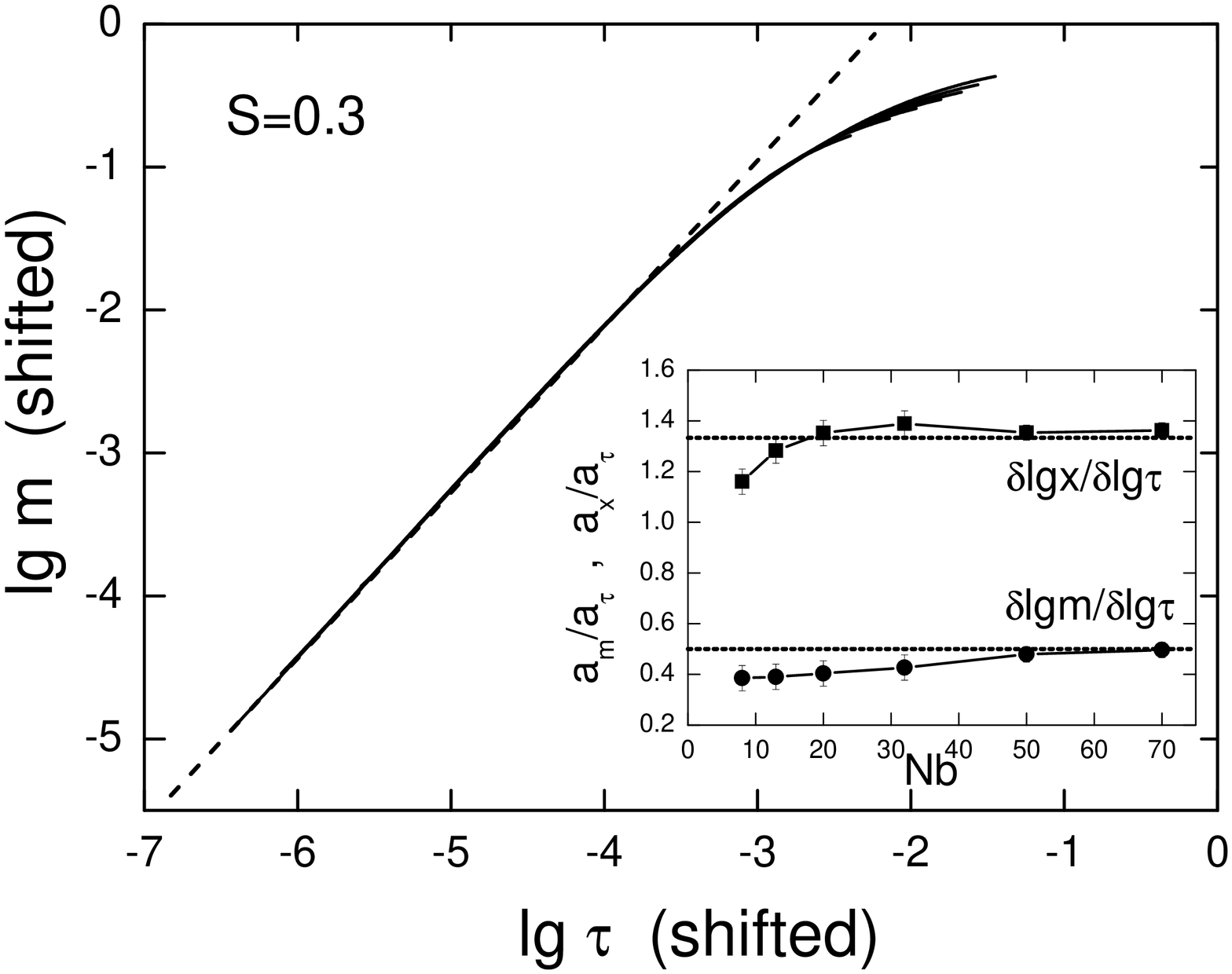}
\vspace*{-1.0cm}
\end{center}
\caption{Data collapse of $\text{lg}m(\epsilon=0)$ vs.
$\text{lg}\tau$ curves shown in Fig.1. The dashed line has a slope
$(1-s)/(2s)=1.167$. Inset: the ratios of the shift magnitudes
$\delta \text{lg}x/\delta \text{lg}\tau$ (squares) and $\delta
\text{lg}m / \delta \text{lg}\tau$ (dots) as functions of $N_{b}$.
The dashed lines are $a_{x}/a_{\tau}=(1-2s)/s=1.333$ (upper) and
$a_{m}/a_{\tau}=1/2$ (lower), respectively.}
\end{figure}
In Fig.1, we show the $N_b$ dependence of the function $m(\tau)$,
with other parameters fixed. We observe that in the small $\tau$
limit, $m(\tau) \propto \tau^{\beta_{NRG}}$ and the slope is
independent of $N_{b}$. To an accuracy of $0.1\%$, the extracted
exponent $\beta_{NRG}$ agrees with $(1-s)/(2s)$. We checked other
$s$ values in the regime $0<s<1/2$ and the fitted $\beta_{NRG}$
fulfill Eq.(17) very well, as shown in Fig.14(a). As discussed in
Appendix, the expressions in Eq.(17) are the results of a specific
parametrization for the bath degrees of freedom, i.e., the
logarithmic discretization. In the regime $0<s<1/2$, different
$\beta$ and $\delta$ will be produced if different parameterization
is used. Thus the coincidence with Eq.(17) hints that $\beta$ and
$\delta$ produced by BNRG are actually artificial ones induced by
the boson state truncation and their values dominated by the
logarithmic discretization. Further evidence of the interplay
between $N_{b}$ and $\Lambda$ is given below.

For a fixed $\text{lg}\tau$, $m(\tau) \propto x^{-1/2}$ in the large
$N_{b}$ limit, as shown in the inset of Fig.1. Therefore, in the
small $\tau$ limit, we have a double power form
\begin{equation}
   m(\tau, \epsilon=0, \Delta, x, w) \propto \tau^{\beta_{NRG}}
   x^{-1/2}.
\end{equation}
As $N_{b}$ increases, the curve shifts along certain directions,
signaling the scaling behavior. The upper part of the curve has an
approximate slope $\beta_{MF}=1/2$. Its range is enlarged as $N_{b}$
increase. These features resemble what was found in the mean-field
spin-boson model.~\cite{Hou1} It is then expected that the crossover
$\tau_{cr}$ between the two power law regimes: the lower one with
$\beta_{NRG}$ and the upper one with $\beta_{MF}$, moves toward zero
as $x$ tends to zero. Thus at $x=0$ the classical exponent
$\beta_{MF}$ will be recovered in the whole $\tau$ regime.

Following Ref.~\onlinecite{Hankey1}, we assume that $m(\tau, x)$ is
a generalized homogeneous function (GHF), i.e.,
$m(\tau\lambda^{a_{\tau}}, x\lambda^{a_{x}}) = \lambda^{a_m} m(\tau,
x)$ for any positive $\lambda$. Letting $\lambda =
\tau^{1/a_{\tau}}$, we get the scaling form
\begin{equation}
 \bar{m}(\bar{\tau}, \bar{x}) = \frac{a_{m}}{a_{\tau}} \bar{\tau}+ \bar{g}( \bar{x}- \frac{a_{x}}{a_{\tau}} \bar{\tau}).
\end{equation}
Here, $\bar{t}\equiv \text{log}_{10} t$ ($t=m, \tau, x$). $a_{\tau}$
and $a_{x}$ are the scaling powers for $\tau$ and $x$, respectively.
$\bar{g}(\bar{z})$ is a universal function. Using information from
Fig.1, i.e., the double power form Eq.(19) in small $\tau$ limit and
the non-sigularity of $m(\tau, x)$ in the limit $x \rightarrow 0$,
it is easy to obtain the following ansatz for $\bar{g}(\bar{z})$,
\begin{eqnarray}\label{eq:24}
\bar{g}(\bar{z}) \propto \left\{
\begin{array}{lll}
 const. ,\,\,\,\,\,   & (\bar{z}_{sat} \ll z \ll \bar{z}_{cr}); \\
&\\
 \theta \bar{z} ,\,\,\,\,\, & (\bar{z} \gg \bar{z}_{cr}).
\end{array} \right.
\end{eqnarray}
Here $\bar{z}=\bar{x}- a_{x}/a_{\tau} \bar{\tau}$. In the saturation
regime where $\bar{z} \ll \bar{z}_{sat}$, $\bar{g} \propto
(a_{m}/a_{x})(\bar{z} - \bar{x})$. From Eqs.(17),(19), and (20), one
extracts the exponents $\theta=-1/2$ and
\begin{eqnarray}
&& a_{m}/a_{\tau}=\beta_{MF}=1/2 \,\,\,\, \nonumber \\
&& a_{x}/a_{\tau}=(1-2s)/s \,\,\,\,.
\end{eqnarray}
One way to verify the scaling ansatz is to show the data collapse.
The GHF assumption implies that
\begin{equation}
   \bar{m}(\bar{\tau}+a_{\tau}\bar{\lambda},
   \bar{x}+a_{x}\bar{\lambda})=\bar{m}(\bar{\tau},
   \bar{x})+a_{m}\bar{\lambda}.
\end{equation}
Therefore, in the log-log diagram, the group of curves $m(\tau,x)$
should collapse when $\bar{\tau}$, $\bar{x}$, and $\bar{m}$ are
shifted by $\delta \bar{\tau}$, $\delta \bar{x}$, and $\delta
\bar{m}$, respectively. The ratios between any two of them give the
corresponding exponents, $\delta \bar{m}/\delta
\bar{\tau}=a_{m}/a_{\tau}$ and $\delta \bar{x}/ \delta
\bar{\tau}=a_{x}/a_{\tau}$. In Fig.2, a perfect data collapse is
obtained from the data in Fig.1. The ratios of the shifts are
plotted as functions of $N_{b}$ in the inset. Compared with Eq.(22),
the agreement is poorer for smaller $N_{b}$, probably due to
nonuniversal corrections. It improves continually as $N_{b}$
increases. This forms a consistent confirmation of the GHF
assumption and the results Eqs.(17),(19)-(22).
\begin{figure}[t!]
\begin{center}
\includegraphics[width=2.7in, height=3.9in, angle=270]{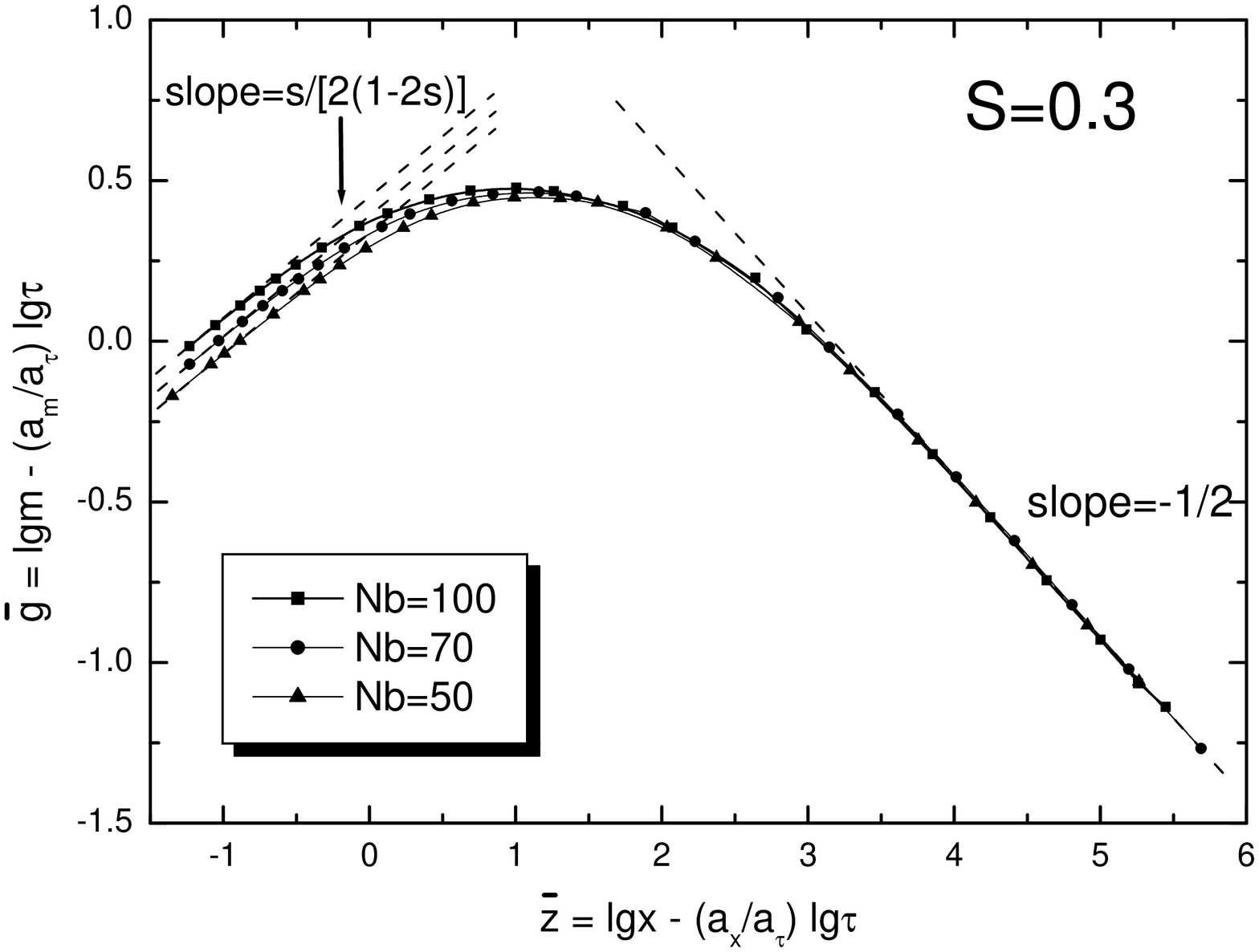}
\vspace*{-1.0cm}
\end{center}
\caption{$\bar{g}=\text{lg}m-a_{m}/a_{\tau}\text{lg}\tau$ vs.
$\bar{z}=\text{lg}x-a_{x}/a_{\tau}\text{lg}\tau$ for various
$N_{b}$'s at $s=0.3, \Delta=0.1$ (squares with solid guiding lines).
Here $a_{m}/a_{\tau}=1/2$ and $a_{x}/a_{\tau}=(1-2s)/s$ are used.
From bottom to top, $N_{b}=50, 70, 100$, respectively. The dashed
lines with slope $s/(2(1-2s))$ and $-1/2$ mark the large and small
$\tau$ limit, respectively. As $N_{b}$ increases, the regime with
almost zero slope expands.}
\end{figure}

The universal function $\bar{g}(\bar{z})$ is plotted in Fig.3 for
$N_{b} \geq 50$. The downturn on the left part of the curve comes
from the saturation of $m$ in the large $\tau$ regime. As $x$ gets
smaller, the intermediate regime with zero slope extends, forming a
pronounced plateau as described by Eq.(21). In the $m(\tau)$ curve,
this corresponds to the extension of the regime with $\beta=1/2$ as
$N_{b}$ increases.

The two-section behavior of $\bar{g}(\bar{z})$ in Eq.(21) agrees
with Eq.(10). Putting Eq.(21) into Eq.(20), one gets $m(\tau, x)
\propto x^{-1/2}\tau^{\beta_{NRG}}$ for $\tau \ll \tau_{cr}$ and
$m(\tau, x) \propto \tau^{\beta_{MF}}$ for $\tau \gg \tau_{cr}$. The
crossover $\tau_{cr}$ is determined by $z_{cr}^{0} =z_{cr}^{\theta}$
and $\bar{z}=\bar{x}-a_{x}/a_{\tau}\bar{\tau}$. We get $\tau_{cr}
\propto x^{a_{\tau}/a_{x}}=x^{s/(1-2s)}$, same as in Eq.(11).
\begin{figure}[t!]
\begin{center}
\includegraphics[width=3.3in, height=4.0in, angle=270]{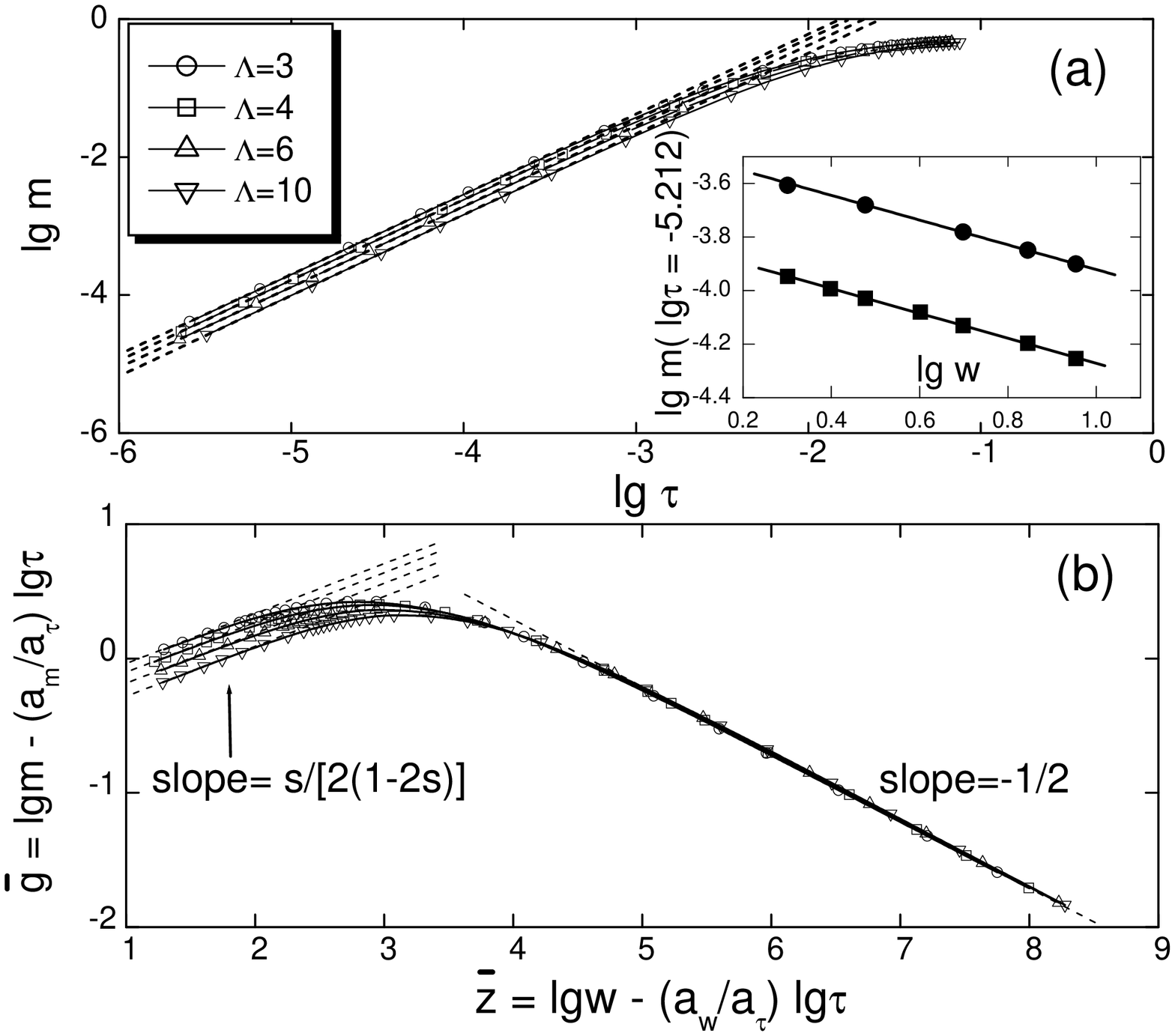}
\vspace*{-1.0cm}
\end{center}
\caption{$\Lambda$ scaling behavior of the order parameter at
$s=0.3, \Delta=0.1$. (a) $\text{lg}m(\epsilon=0)$ vs.
$\text{lg}\tau$ for various $\Lambda$'s ($N_{b}=8, M_{s}=80$).
Symbols with guiding lines are BNRG data and dashed lines are fitted
lines. Inset: $\text{lg}m(\epsilon=0)$ at $\text{lg}\tau=-5.212$ as
functions of $\text{lg}w$ for $N_{b}=8, M_{s}=80$ (squares) and
$N_{b}=20, M_{s}=120$ (dots), respectively. The slopes of the
fitting solid lines are $-0.46$. Other parameters are the same as in
Fig.1. (b) Universal function of
$\bar{g}=\text{lg}m-a_{m}/a_{\tau}\text{lg}\tau$ v.s.
$\bar{z}=\text{lg}w-a_{w}/a_{\tau}\text{lg}\tau$ for different
$\Lambda$'s. Here $a_{m}/a_{\tau}=1/2$ and $a_{w}/a_{\tau}=(1-2s)/s$
are used. The dashed lines with slope $s/(2(1-2s))$ and $-1/2$ mark
the large and small $\tau$ limit, respectively. As $w$ decreases,
the regime with almost zero slope expands.}
\end{figure}

Guided by the mean-field results Eq.(10), we also carry out scaling
analysis for $m(\tau, w)$ with respect to $w=\Lambda-1$ (for $s=0.3$
and fixed $N_{b}$). In Fig.4(a), similar scaling behavior as in
$m(\tau, x)$ is observed. In the inset, $\text{lg}m$ is plotted as a
function of $\text{lg}w$ for a fixed $\text{lg}\tau$, giving a power
law behavior with exponent $-0.46$, consistent with the $-1/2$ in
Eq.(10) within numerical errors. By assuming that $m(\tau, w)$ is a
GHF with scaling exponent $(a_{m}, a_{\tau},a_{w})$, using the
exponents Eq.(22), and by applying the data collapse procedure (not
shown), we obtain $a_{m}/a_{\tau}=1/2$ and
$a_{w}/a_{\tau}=(1-2s)/s=a_{x}/a_{\tau}$. In Fig.4(b), the universal
function concerning $m(\tau, w)$ is plotted using the above
exponents. Similar to Fig.3, the universal function has the form of
Eq.(21), with $x \rightarrow w$. Fig.4 shows that the scaling
variable $w$ plays a similar role as $x$. Therefore, the crossover
 scale $\tau_{cr}$ has an additional factor $w^{s/(1-2s)}$. For
either $w=0$ or $x=0$, the $m(\tau, x, w)$ curve will have the
classical exponent $\beta_{MF}$ in the regime $\tau \ll \tau_{sat}$.
Here, again the NRG data agree with the mean-field expressions (10)
and (11). The fact that $w$ becomes a scaling variable means the
failure of the logarithmic discretization in NRG: $\Lambda$ may
alter the universal quantities. This is solely due to the boson
state truncation. As learned from the mean-field study, the
parametrization scheme for the bath is relevant for the critical
behavior once the bosons are no longer canonical.

\begin{figure}[t!]
\begin{center}
\includegraphics[width=3.3in, height=4.1in, angle=270]{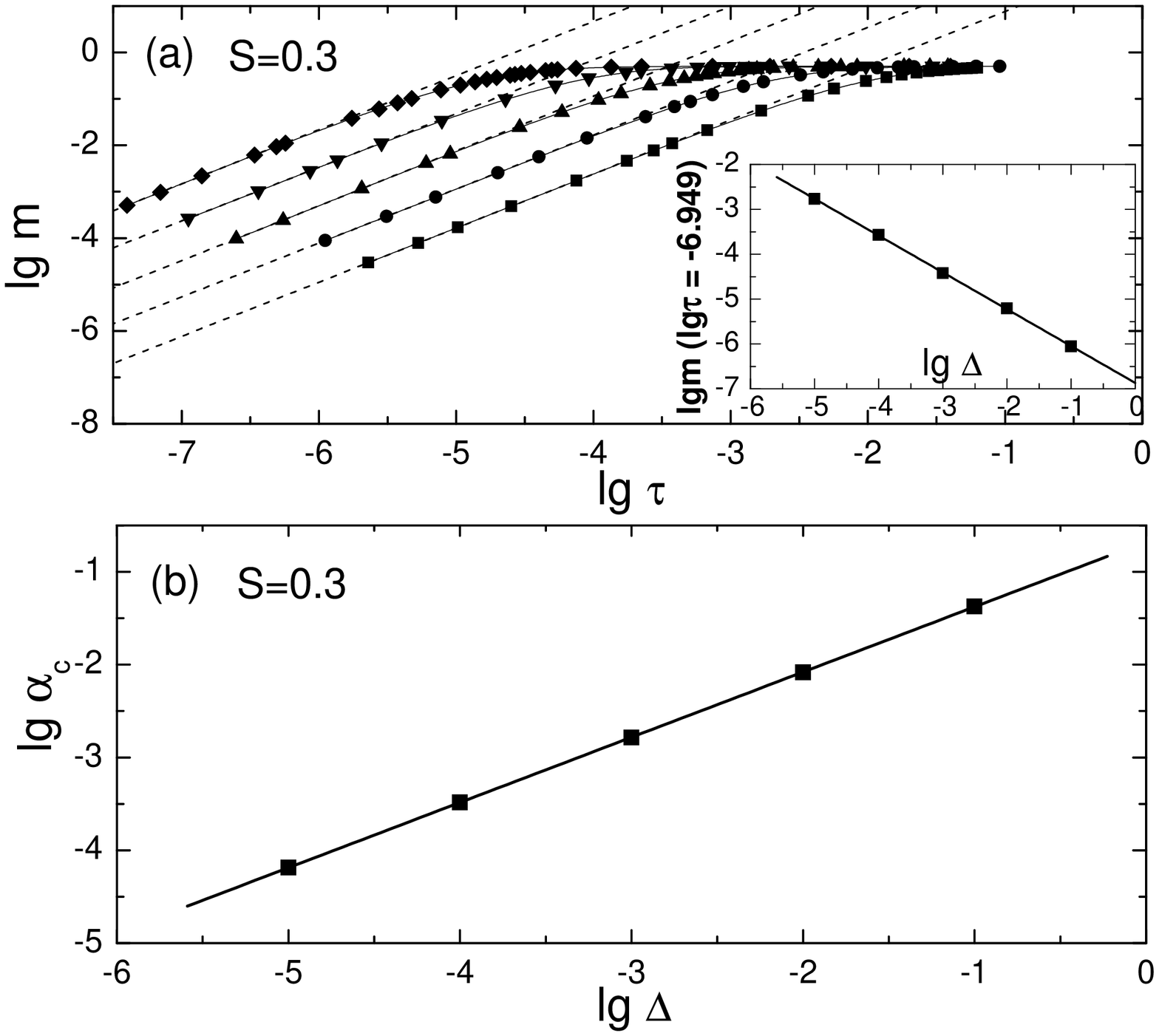}
\vspace*{-0.7cm}
\end{center}
\caption{(a) $\text{lg}m(\epsilon=0)$ vs. $\text{lg}\tau$ for
different $\Delta$'s at $s=0.3$. From bottom to top,
$\Delta=10^{-1}, 10^{-2},10^{-3},10^{-4},10^{-5}$, respectively.
Symbols are BNRG data and the dashed lines are power law fit from
small $\tau$ data, which agree with $\beta_{NRG}=(1-s)/(2s)$. Inset:
$\text{lg}m$ at $\text{lg}\tau=-16$ as a function of $\Delta$
(symbols). The fitted line (solid line) has a slope $-0.821$ which
agrees well with $-(1-s)^{2}/(2s)=-0.817$. (b)$\text{lg}\alpha_c$ as
a function of $\text{lg}\Delta$. Symbols are BNRG data and the solid
line is the fitted line with slope 0.703 which agrees with $1-s$.
Calculated with $N_b=8$, $M_s=80$, and $\Lambda=4.0$.}
\end{figure}

As for $\Delta$, we show the BNRG results in Fig.5. The power law
fits in Fig.5(a) and its inset disclose a double power behavior in
the small $\tau$ limit: $m(\tau, \Delta) \propto
\Delta^{\eta}\tau^{\beta_{NRG}}$. Here $\eta$ is an exponent to be
determined below. In Fig.5(b), we show that BNRG produces
$\alpha_c(1) \propto \Delta^{1-s}$, a result that has been
established by BNRG and perturbative renormalization group
study.~\cite{Vojta1}
 To understand the exponent $\eta$,
we resort to Eq.(10). If we collect the exponent of $\Delta$ in
Eq.(10), and are careful enough to use the BNRG result
$\alpha_c(1)=c(\Delta/\omega_c)^{1-s}$ (instead of the mean-field
one $\alpha_c(1)=\Delta s/(2\omega_c)$) in Eq.(10), we get
$\eta=-(1-s)^{2}/(2s)=-0.817$ for $s=0.3$. The exponent fitted from
BNRG data is $-0.821$, as shown in the inset of Fig.5(a). The
excellent agreement supports that the $\Delta$-dependence in BNRG
results can also be summarized by Eq.(10). We checked other $s$
values in $0<s<1$ and confirmed this conclusion.

Combining the above results, we draw the conclusion that in the
regime $0<s<1/2$, BNRG resutls for $m(\tau, \epsilon=0, \Delta,
w,x)$ are well described by Eqs.(10), except that the
$\Delta$-dependence of $\alpha_c(1)$ in the equation should be
replaced by the BNRG form Eq.(18). Correspondingly, the crossover
point $\tau_{cr}$ reads as

\begin{equation}
   \tau_{cr} \propto \Delta^{1-s} (wx)^{s/(1-2s)}.
\end{equation}

As a scaling variable, $\Delta$ is different from $w$ or $x$. As
seen in Fig.5(a), when $\Delta$ decreases, both the $\beta_{NRG}$-
and $\beta_{MF}$-exponent regimes in $m(\tau)$ curve shift to the
left along the horizontal direction, but the $\beta_{MF}$-exponent
regime does not expand. This is because another crossover scale
$\tau_{sat}$, above which $m(\tau) \sim 1/2$, also decreases as
$\Delta^{1-s}$. Indeed, using $m(\tau, \Delta) \propto [\tau/
\alpha_c(1) ]^{1/2} \propto (\tau/\Delta^{1-s})^{1/2}$ in $\tau_{cr}
\ll \tau \ll \tau_{sat}$ and $m(\tau, \Delta) \sim 1/2$ in $\tau \gg
\tau_{sat}$, one gets $\tau_{sat} \propto \Delta^{1-s}$, same as
$\tau_{cr}$.

\subsubsection{ $ m(\tau=0, \epsilon, \Delta, x, w) $ }
In this part, we fix $\tau=0$ and study the order parameter $m$ as a
function of $\epsilon$. This is related to the exponent $\delta$.

\begin{figure}[t!]
\begin{center}
\includegraphics[width=2.6in, height=3.7in, angle=270]{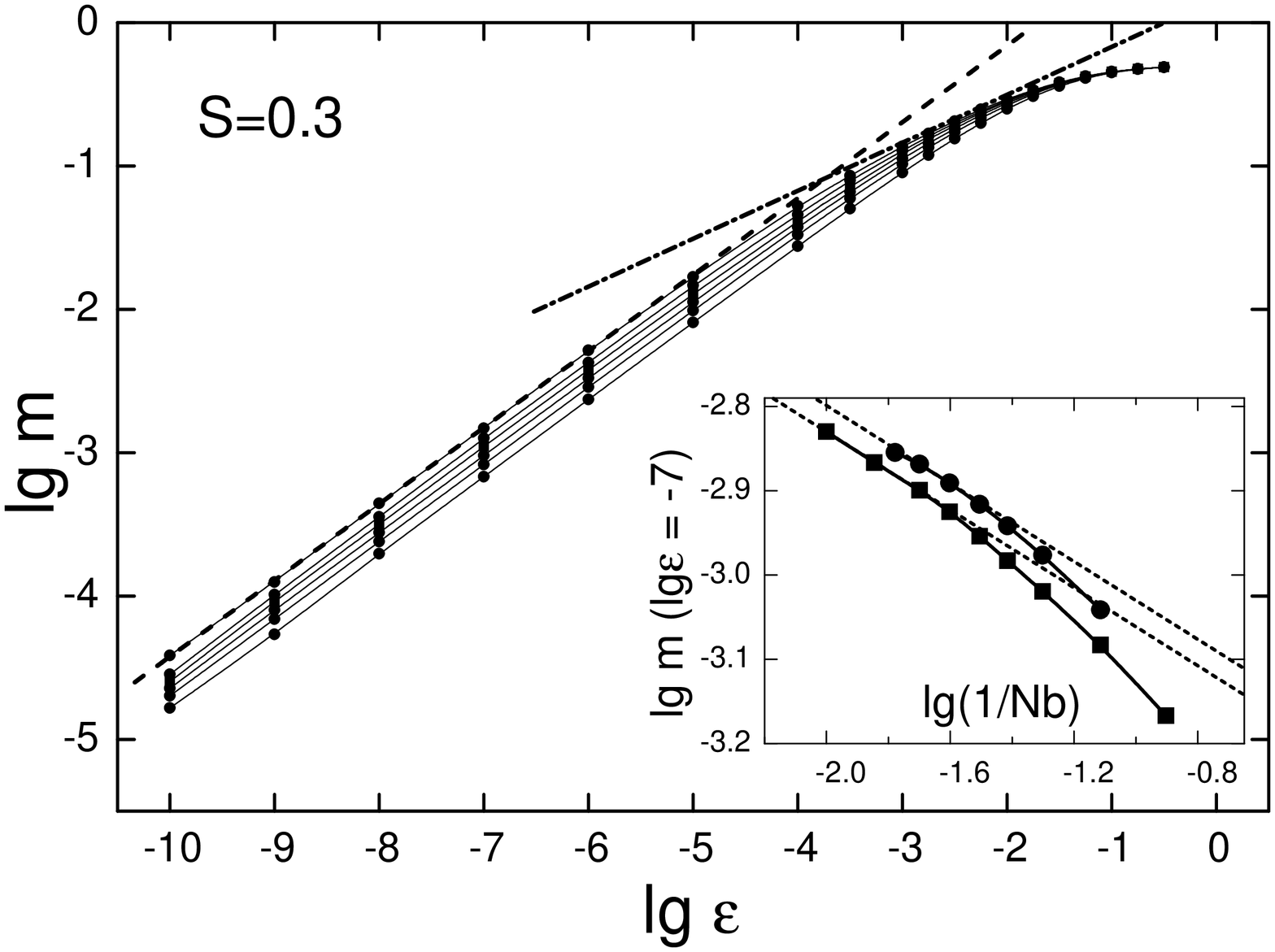}
\vspace*{-1.0cm}
\end{center}
\caption{$\text{lg}m(\tau=0)$ vs. $\text{lg}\epsilon$ for different
$N_{b}$'s at $s=0.3, \Delta=0.1$. From bottom to top, $N_{b}=8, 13,
20, 32, 50, 100$, respectively. Other parameters are $\Delta=0.1$,
$\Lambda=10$, $M_{s}=80$. Symbols with solid guiding lines are BNRG
data. The dashed line is power law fit for $N_{b}=100$ in small
$\epsilon$ limit, which gives slope $1/\delta_{NRG}=0.533$ being
consistent with $(1-s)/(1+s)$. The guiding dash-dotted line marks
 the asymptotic behavior $1/\delta=1/3$ in the large $N_{b}$ limit. Inset:
$\text{lg}m(\tau=0)$ at $\epsilon=10^{-7}$ as functions of
$\text{lg}x$ for $\Lambda=10, M_{s}=80$ (squares) and $\Lambda=4,
M_{s}=120$ (dots), respectively. The dashed lines are guiding lines
with slope $-s/(1+s)=-0.231$ to mark the power law behavior in large
$N_{b}$ limit.}
\end{figure}

In Fig.6, $m(\epsilon, x)$ versus $\epsilon$ at $\tau=0$ is plotted
for various $x$ values. Similar to $m(\tau, x)$, in the
small-$\epsilon$ limit all curves for different $N_{b}$'s have the
same power law $m(\epsilon, x) \propto \epsilon^{1/\delta_{NRG}}$
with $x$-dependent coefficients. For $s=0.3$, the fitted power
$0.533$ agrees well with the mean-field expression
$(1-s)/(1+s)=0.538$. For other $s$ values in $0<s<1/2$, the
comparison is shown in Fig.14(b). The inset of Fig.6 supports
$m(\epsilon,x) \propto x^{\theta^{\prime}}$ for fixed $\epsilon$ and
large $N_{b}$, with the exponent $\theta^{\prime}$ to be determined
below. Therefore, a double power form in the small $\epsilon$ limit
is obtained, $m(\epsilon, x) \propto x^{\theta^{\prime}}
\epsilon^{(1-s)/(1+s)}$. In the other limit where $\epsilon$ is much
larger, the upper dashed line in Fig.6 marks out a finite regime
where $m(\epsilon, x) \propto
\epsilon^{1/\delta_{MF}}=\epsilon^{1/3}$. There is a crossover
$\epsilon_{cr} \propto
x^{-\theta^{\prime}/(1/\delta_{NRG}-1/\delta_{MF})}$ which separates
the lower double-power regime from the upper classical regime.
$\epsilon_{cr}$ goes to zero as $x \rightarrow 0$.

To obtain $\theta^{\prime}$, we carry out the scaling
analysis based on GHF and do the data collapse in Fig.7. We assume
that $m(\epsilon, x)$ is a GHF with scaling powers $(a_{m},
a_{\epsilon}, a_{x})$ and get
$m(\epsilon,x)=\epsilon^{a_m/a_{\epsilon}}h(x/\epsilon^{a_{x}/a_{\epsilon}})$.
The universal function $h(z)$ has a two-regime form similar to
Eq.(21), i.e.,

\begin{eqnarray}
\bar{h}(\bar{z}) \propto \left\{
\begin{array}{lll}
 const. ,\,\,\,\,\,   & (\bar{z}_{sat} \ll \bar{z} \ll \bar{z}_{cr}); \\
&\\
 \theta^{\prime} \bar{z} ,\,\,\,\,\, & (\bar{z} \gg \bar{z}_{cr}).
\end{array} \right.
\end{eqnarray}
Here $\bar{z}=\bar{x}-(a_{x}/a_{\epsilon})\bar{\epsilon}$. Using
$m(\epsilon, 0) \propto \epsilon^{1/3}$ and
$a_{m}/a_{x}=s/[2(1-2s)]$ confirmed before, we obtain
\begin{eqnarray}
 m(\epsilon, x) \propto \left\{
\begin{array}{lll}
  x^{-s/(1+s)}\epsilon^{1/\delta_{NRG}},  \,\,\,\,\,   & (\epsilon \ll \epsilon_{cr}); \\
&\\
 \epsilon^{1/3}     \,\,\,\,\, & (\epsilon \gg \epsilon_{cr}).
\end{array} \right.
\end{eqnarray}
Here $\epsilon_{cr} \propto x^{a_{\epsilon}/a_{x}}$. This means
$\theta^{\prime} =-s/(1+s)$ and
\begin{eqnarray}
  && \frac{a_{m}}{a_{\epsilon}}=\frac{1}{3}, \,\,\,\,\,  \nonumber \\
  && \frac{a_x}{a_{\epsilon}}=\frac{2(1-2s)}{3s}.
\end{eqnarray}
In the inset of Fig.6, the BNRG result for $\theta^{\prime}$
(symbols) compares favorably with the exponent $-s/(1+s)$ (dashed
lines). In the inset of Fig.7, the
exponents $a_{m}/a_{\epsilon}$ and $a_{x}/a_{\epsilon}$ extracted
from ratios of the shifts $\delta m$, $\delta \epsilon$, and $\delta
x$ agree well with Eq.(27) in the large $N_b$ limit, confirming
Eq.(26). The similarity to the mean-field equation Eq.(14) is obvious.
 Using this $\theta^{\prime}$ we also obtain $\epsilon_{cr} \propto
x^{3s/[2(1-2s)]}$. It agrees with the mean-field result in Eq.(15).
\begin{figure}[t!]
\begin{center}
\includegraphics[width=2.5in, height=3.7in, angle=270]{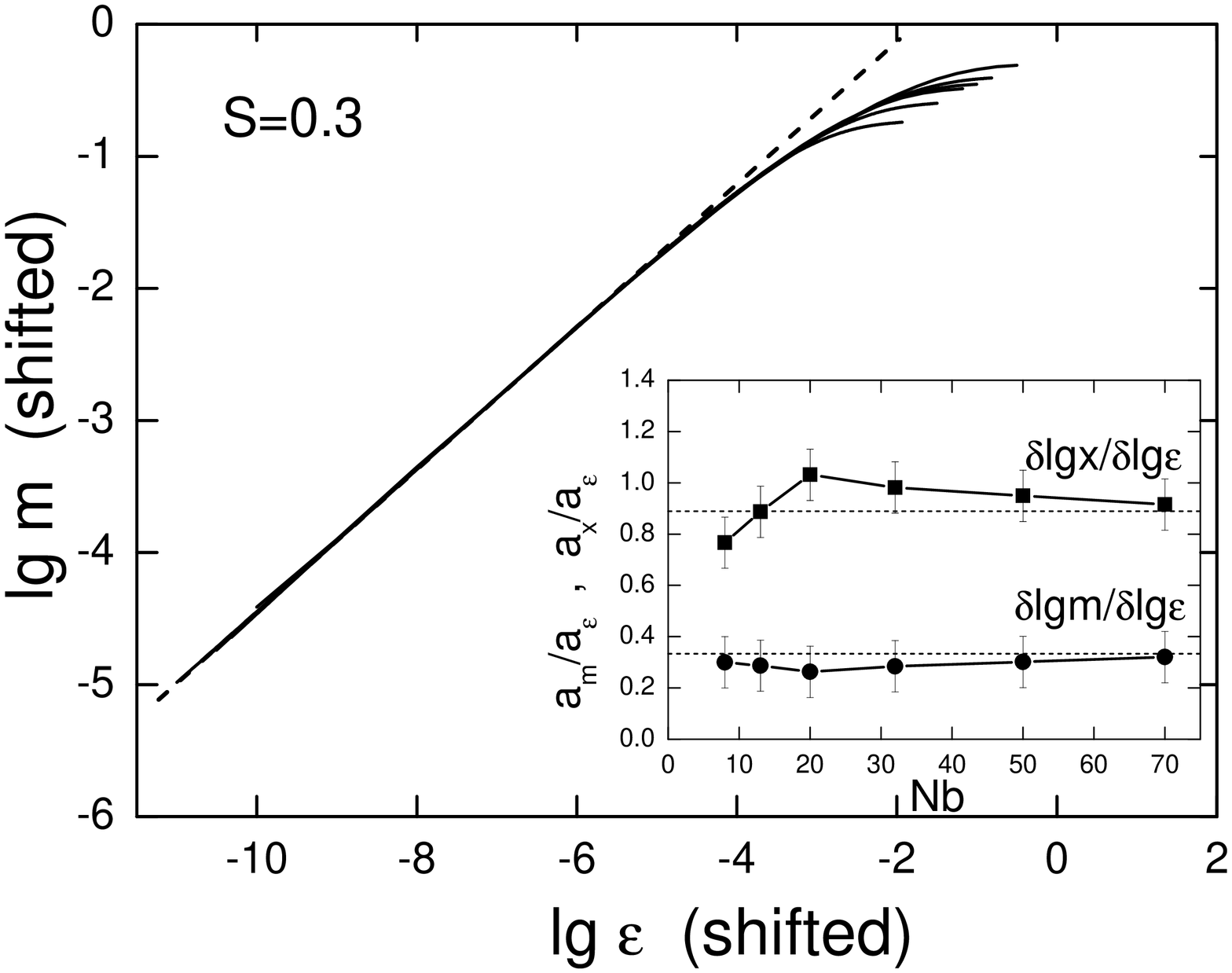}
\vspace*{-0.6cm}
\end{center}
\caption{Data collapse of $\text{lg}m(\tau=0)$ vs.
$\text{lg}\epsilon$ curves shown in Fig.6. The dashed straight line
has a slope $(1-s)/(1+s)=0.538$. Inset: the ratio of the shift
magnitudes $\delta \text{lg}x/\delta \text{lg}\epsilon$ (squares),
and $\delta \text{lg}m / \delta \text{lg}\epsilon$ (dots) as
functions of $N_{b}$. The dashed straight lines are
$a_{x}/a_{\epsilon}=2(1-2s)/(3s)=0.889$ (upper) and
$a_{m}/a_{\epsilon}=1/3$ (lower), respectively. }
\end{figure}
\begin{figure}[t!]
\begin{center}
\includegraphics[width=3.4in, height=3.8in, angle=270]{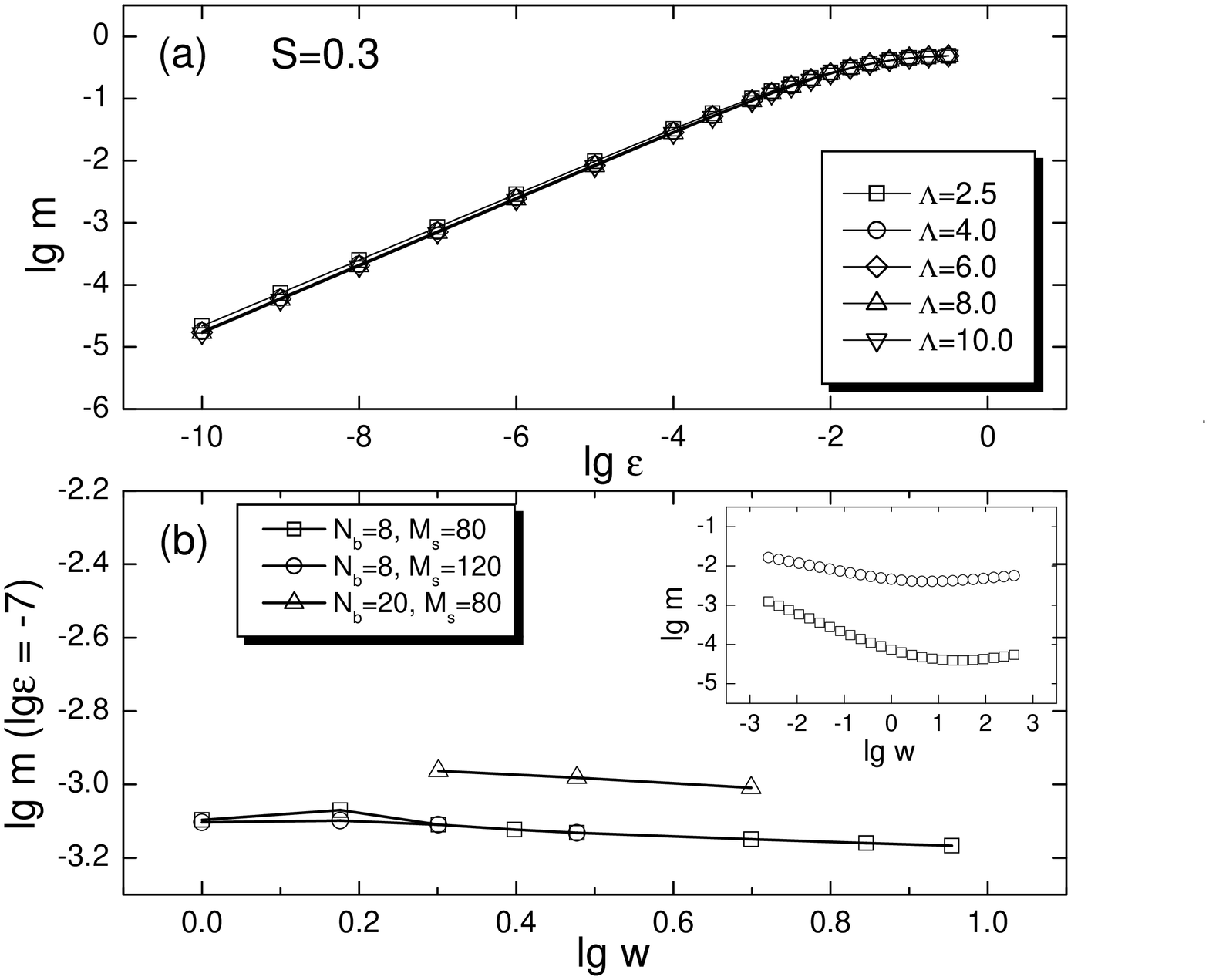}
\vspace*{-0.5cm}
\end{center}
\caption{(a)$\text{lg}m(\tau=0)$ vs. $\text{lg}\epsilon$ for
different $\Lambda$'s at $s=0.3$. (b)$\text{lg}m(\tau=0)$ vs.
$\text{lg}w$ at $\epsilon=10^{-7}$ for various ($N_{b},M_{s}$)'s.
The fitted slope is about $-0.07$ in the shown regime, different
from $-s/(1+s)=-0.231$. This is probably due to the fact that $w$ is
not small enough. Inset: the mean-field results $\text{lg}m$ vs.
$\text{lg}w$ for $\tau=0,\epsilon=10^{-7}$ (cycles) and
$\tau=10^{-7}$,$\epsilon=0$ (squares), respectively. Other
parameters are $\Delta=0.1$, $N_{b}=8$. }
\end{figure}

The $\Lambda$ scaling analysis is carried out in Fig.8 for
$m(\tau=0,\epsilon, w)$ for fixed $\Delta$, $N_{b}$ and $M_{s}$. In
Fig.8(a), the small $\epsilon$ regime of the curves fulfills the
power law $m(\epsilon, w) \propto \epsilon^{1/\delta_{NRG}}$ and the
larger $\epsilon$ regime $m(\epsilon, w) \propto
\epsilon^{1/\delta_{MF}}$. However, for a fixed $\epsilon$, the
shift of the curve with $\Lambda$ is very weak for $\Lambda$ between
$2.5$ and $10$. In Fig.8(b), the order parameter
$m(\epsilon=10^{-7}, w)$ does not fit a power law of $w$ as nicely
as $m(\tau=\tau_{0},w)$ does in Fig.4. The data in the small
$\Lambda$ regime are $M_{s}$ dependent. Here the crudest estimation
from larger $\Lambda$ regime gives the power $-0.07$. Assuming
a GHF form for $m(\epsilon,w)$ (at fixed $\tau=0$ and $x >0$) and
using previously obtained relations among $a_{m}, a_{\tau},
a_{\epsilon}$ and $a_{w}$, we get
\begin{equation}
  m(\epsilon, w) \propto \epsilon^{1/\delta_{NRG}} w^{-s/(1+s)},
\end{equation}
same as in the mean-field result Eq.(14). The fitted power $-0.07$ is quite
different from $-s/(1+s)=-0.231$.

We believe that this discrepancy is due to the fact that for $2
\leqslant \Lambda \leqslant 10$, $w$ is not small enough to enter
the scaling regime for observing the correct power law. This is in
contrast to Fig.4 where a nice $w$-scaling persists up to
$\Lambda=10$. We have not fully understood this contrast yet, but
only mention that we observed similar differences in the mean-field
calculations. There, the shift with $w$ is much more pronounced in
$m(\tau=\tau_0,\epsilon=0,w)$ than in
$m(\tau=0,\epsilon=\epsilon_0,w)$. The curves
$\text{lg}m-\text{lg}w$ of the mean-field Hamiltonian are plotted in
the inset of Fig.8(b), for $\tau=0$ (cycles) and $\epsilon=0$
(squares), respectively. It is seen that the expected power law
behavior appears only when $\Lambda \ll 2$. In the regime $2 <
\Lambda < 10$, the $\text{lg}m(\tau=0,\epsilon=\epsilon_0,w)$ curve
significantly deviates from the correct power law behavior,
resembling the BNRG results. This supports our notion that much
smaller $\Lambda$ is required to observe the
$m(\tau=0,\epsilon=\epsilon_{0},w) \propto w^{-s/(1+s)}$ behavior.
As a consequence of this reasoning, $w$ should also appear in the
crossover scale $\epsilon_{cr}$ as a factor $w^{3s/(2(1-s))}$.

We checked our results for $N_{b}$ and $M_{s}$ up to
$50$ and $300$, respectively, and found that the quality of the data
is not improved. In the BNRG calculations, after each diagonalization,
 only the lowest $1/N_{b}$ fraction of eigen-states
are kept. Hence, as $N_{b}$ increases, one needs larger $\Lambda$ or
larger $M_{s}$ to compensate the error from discarding states. As a
result, it is very difficult to produce reliable data in the
large-$N_{b}$ and small-$\Lambda$ regimes. For a fixed $N_{b}$, it
is known that to approach a smaller $\Lambda$ regime, one needs to
use larger $M_{s}$. However, we did not find the expected power law
behavior of $w$ up to $\Lambda=2$ using $N_{b}=8$ and $M_{s}=300$.

\begin{figure}[t!]
\begin{center}
\includegraphics[width=2.7in, height=3.7in, angle=270]{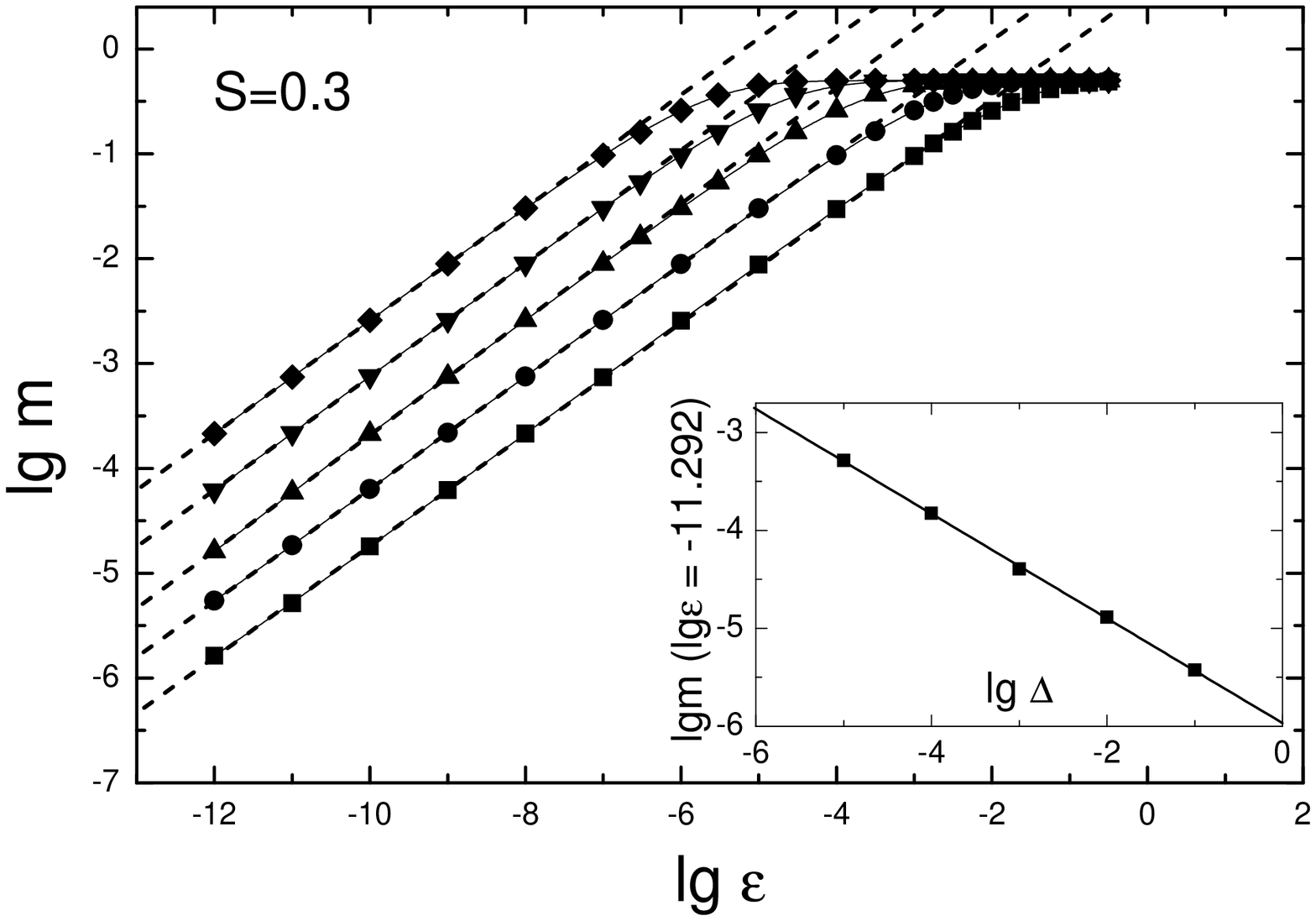}
\vspace*{-2.0cm}
\end{center}
\caption{$\text{lg}m(\tau=0)$ vs. $\text{lg}\epsilon$ for different
$\Delta$'s at $s=0.3$. From bottom to top, $\Delta=10^{-1}, 10^{-2},
10^{-3}, 10^{-4}$, and $10^{-5}$, respectively. Symbols are BNRG
data and dashed lines are power law fit, giving an average slope
$0.539$ which agrees with $(1-s)/(1+s)=0.538$. Inset:
$\text{lg}m(\tau=0)$ at $\text{lg}\epsilon=-11.292$ as a function of
$\text{lg} \Delta$ (symbols). The fitting (solid line) gives a slope
$-0.534$, being consistent with $\epsilon/\Delta$ scaling. Other
parameters are $N_b=8$, and $M_s=80$.}
\end{figure}

In Fig.9, we study the $\Delta$ scaling in $m(\epsilon, \Delta)$ at
$\tau=0$. The figure resembles that of
$\text{lg}m(\tau,\Delta)-\text{lg}\Delta$. The double power form in
the small $\epsilon$ regime is found to fulfill the
$\epsilon/\Delta$ scaling, i.e.,
\begin{equation}
    m(\epsilon,\Delta) \propto
    \left( \frac{\epsilon}{\Delta} \right)^{1/\delta_{NRG}},
\end{equation}
with $1/\delta_{NRG}=(1-s)/(1+s)$. As $\Delta$ decreases, both
$1/\delta_{MF}$- and $1/\delta_{NRG}$-exponent regimes move toward
smaller $\epsilon$ along the horizontal direction. Simple analysis
shows that a crossover scale $\epsilon_{sat} \propto \Delta$
separates the saturation regime $m \sim 1/2$ from the
$1/\delta_{MF}$-exponent regime. $\epsilon_{cr}$ must have the same
factor $\Delta$ because the classical exponent regime is not
enlarged as $\Delta$ deceases. Summarizing these results, we get
$\epsilon_{sat}
> \epsilon_{cr}$ and
\begin{eqnarray}
  && \epsilon_{cr} \propto (wx)^{\frac{3s}{2(1-2s)}}\Delta
                 \,\,\,\,\, , \nonumber \\
  && \epsilon_{sat} \propto \Delta \,\,\,\,\, .
\end{eqnarray}
These results concerning the scaling behavior with $\Delta$ is also
consistent with the mean-field expression Eq.(14)-(15), provided
that we use the BNRG result $\alpha_{c}(1) \propto \Delta^{1-s}$ in
the equations.

\subsubsection{Summary for $0< s < 1/2$}
We summarize the above analysis. Our main conclusion is that in the
regime $0<s<1/2$, due to the boson state truncation, the order
parameter $m$ produced by BNRG is a scaling function of variables
$\tau$, $\epsilon$, $\Delta$, $x=1/N_{b}$, and $w=\Lambda-1$.
Interestingly, this function agrees well with the mean-field
equations for finite $N_{b}$, [Eqs.(10) and (11) and (14) and (15)],
except that the $\Delta$-dependence of $\alpha_c(1)$ should be
replaced by the corresponding BNRG one, i.e., $\alpha_c(1) \propto
\Delta^{1-s}$. In the small-$\tau$ or -$\epsilon$ limits, the
boson-state truncation $N_{b}$ introduces artificial exponents
$\beta$ and $\delta$, which are different from the correct classical
values. We summarize the BNRG results in $0<s<1/2$ as the following:

\begin{eqnarray}\label{eq:24}
&&m(\tau, \epsilon=0, \Delta, x, w)   \nonumber \\
&& = \left\{
\begin{array}{lll}
c\left(\frac{\Delta}{\omega_c}
\right)^{-\frac{(1-s)^2}{2s}}(wx)^{-\frac{1}{2}}\tau^{\frac{1-s}{2s}} ,\,   & (\tau \ll \tau_{cr}); \\
&\\
 c^{\prime} \Delta^{-\frac{1-s}{2}} \tau^{\frac{1}{2}} ,\, & ( \tau_{cr} \ll \tau \ll
 \tau_{sat}); \\
& \\
 1/2  ,\, & ( \tau \gg \tau_{sat}),
\end{array} \right. \nonumber \\
&&
\end{eqnarray}
with the two crossover scales $\tau_{sat}> \tau_{cr}$ given by
\begin{eqnarray}
  && \tau_{cr} \propto \Delta^{1-s}(wx)^{\frac{s}{1-2s}}
                 \,\,\,\,\, , \nonumber \\
  && \tau_{sat} \propto \Delta^{1-s} \,\,\,\,\, .
\end{eqnarray}
For $m(\tau=0, \epsilon, \Delta, x, w)$ we have

\begin{eqnarray}
&&m(\tau=0, \epsilon, x, w)   \nonumber \\
&& = \left\{
\begin{array}{lll}
- c \Delta^{-\frac{1-s}{1+s}} (wx)^{-\frac{s}{1+s}}
(\frac{\epsilon}{\omega_c})^{\frac{1-s}{1+s}}
  ,\,\,\,   & (\epsilon \ll \epsilon_{cr} ); \\
&\\
 -c^{\prime} (\frac{\epsilon}{\Delta})^{1/3} ,\,\,\, &(\epsilon_{cr} \ll \epsilon \ll
 \epsilon_{sat}); \\
& \\
 -1/2 ,\,\,\,  &(\epsilon \gg \epsilon_{sat}).
\end{array} \right. \nonumber \\
&&
\end{eqnarray}
with the two crossover scales $\epsilon_{sat} > \epsilon_{cr}$ given
by
\begin{eqnarray}
  && \epsilon_{cr} \propto (wx)^{\frac{3s}{2(1-2s)}}\Delta
                 \,\,\,\,\, , \nonumber \\
  && \epsilon_{sat} \propto \Delta \,\,\,\,\, .
\end{eqnarray}

For the role of $\Delta$ in the order parameter $m$, we observed
$\tau/\Delta^{1-s}$ scaling in the function $m(\tau, \epsilon=0,
\Delta, w, x)$, while the function $m(\tau=0, \epsilon, \Delta, w,
x)$ has $\epsilon/\Delta$ scaling. These features are independent of
the boson state truncation, and hence should hold also in the regime
$1/2<s<1$. As we will see in the next section, this is indeed the
case.

\subsection{BNRG results for $s=0.7$}

\begin{figure}[t!]
\begin{center}
\includegraphics[width=3.2in, height=4.0in, angle=270]{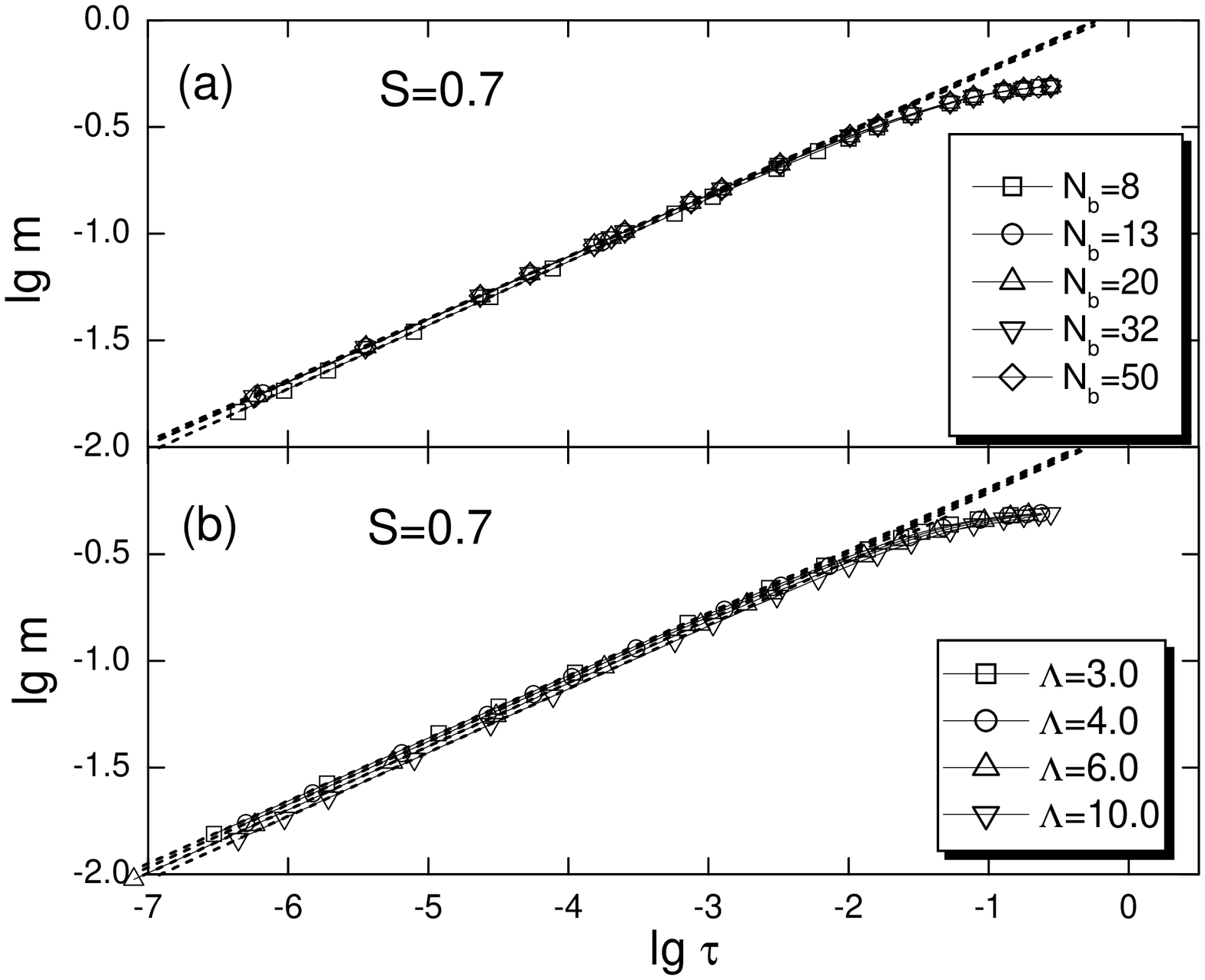}
\vspace*{-1.0cm}
\end{center}
\caption{$\text{lg}m(\epsilon=0)$ vs. $\text{lg}\tau$ at $s=0.7$.
(a) $\Lambda=10.0$ and $N_{b}=8,13, 20,32,50$,respectively; (b)
$N_{b}=8$ and $\Lambda=3.0,4.0,6.0,10.0$, respectively. Symbols are
BNRG data and the dashed lines are power law fits. Other parameters
are $\Delta=0.1$, and $M_s=80$.}
\end{figure}

In this section, to compare with the $0<s<1/2$ case, we study
$s=0.7$, a generic value in the regime $1/2<s<1$. In this regime,
the mean-field theory predicts classical exponents $\beta=1/2$ and
$\delta=3$ for any $N_{b}$. The boson state truncation does not
influence the Gaussian critical fixed point in the mean-field
Hamiltonian.

For the full spin-boson model, BNRG predicts an interacting critical
fixed point and nonclassical exponents $\beta$ and $\delta$. In
Figs. 10(a) and 10(b), $\text{lg}m(\tau, \epsilon=0)-\text{lg}\tau$
curves are plotted for various $N_{b}$'s ($\Lambda=10$) and
$\Lambda$'s ($N_{b}=8$), respectively. Other parameters are fixed.
It is clearly seen that the situation is dramatically different from
$s=0.3$: there is no $N_{b}$ or $\Lambda$ scaling. Therefore, the
boson state truncation and $\Lambda$ do not influence the correct
extraction of $\beta$.

\begin{figure}[t!]
\begin{center}
\includegraphics[width=2.6in, height=3.5in, angle=270]{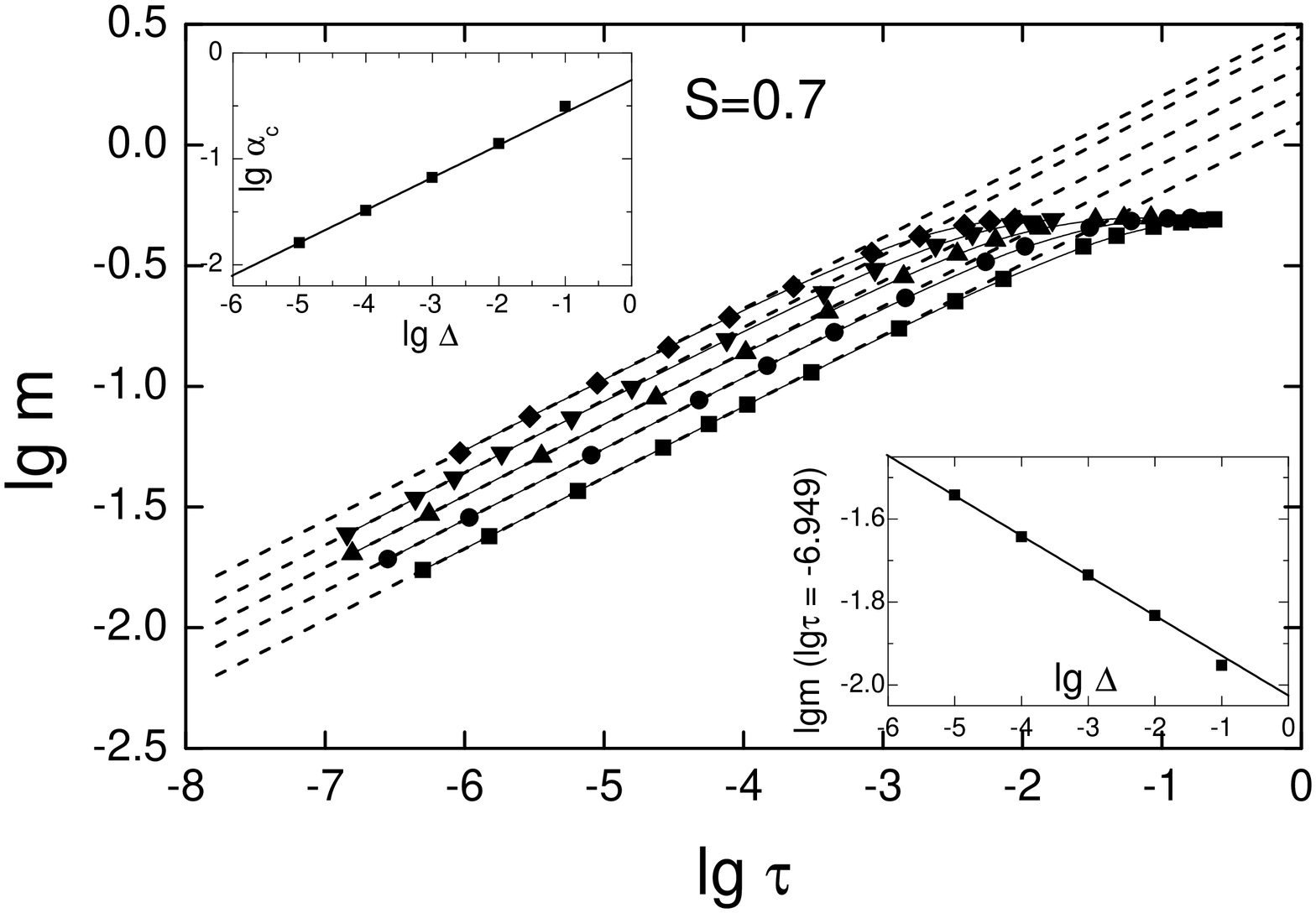}
\vspace*{-1.0cm}
\end{center}
\caption{$\text{lg}m(\epsilon=0)$ vs. $\text{lg}\tau$ at $s=0.7$ for
different $\Delta$'s. From bottom to top, $\Delta=10^{-1},10^{-2},
10^{-3},10^{-4},10^{-5}$, respectively. Symbols are BNRG data and
the dashed lines are power law fit, giving an average slope 0.296.
Inset on top left: $\text{lg}\alpha_c$ vs. $\text{lg}\Delta$ data
gives the fitted slope 0.307, consistent with $1-s$. Inset on bottom
right: $\text{lg}m$ at $\text{lg}\tau=-6.949$ as a function of
$\text{lg}\Delta$ (symbols). The fitted slope is $-0.096$, close to
$-\beta(1-s)=-0.089$, as expected from $\tau/\Delta^{1-s}$ scaling.
Other parameters are $\Lambda=4.0$, $N_{b}=8$, $M_{s}=80$.}
\end{figure}
\begin{figure}[t!]
\begin{center}
\includegraphics[width=3.2in, height=4.0in, angle=270]{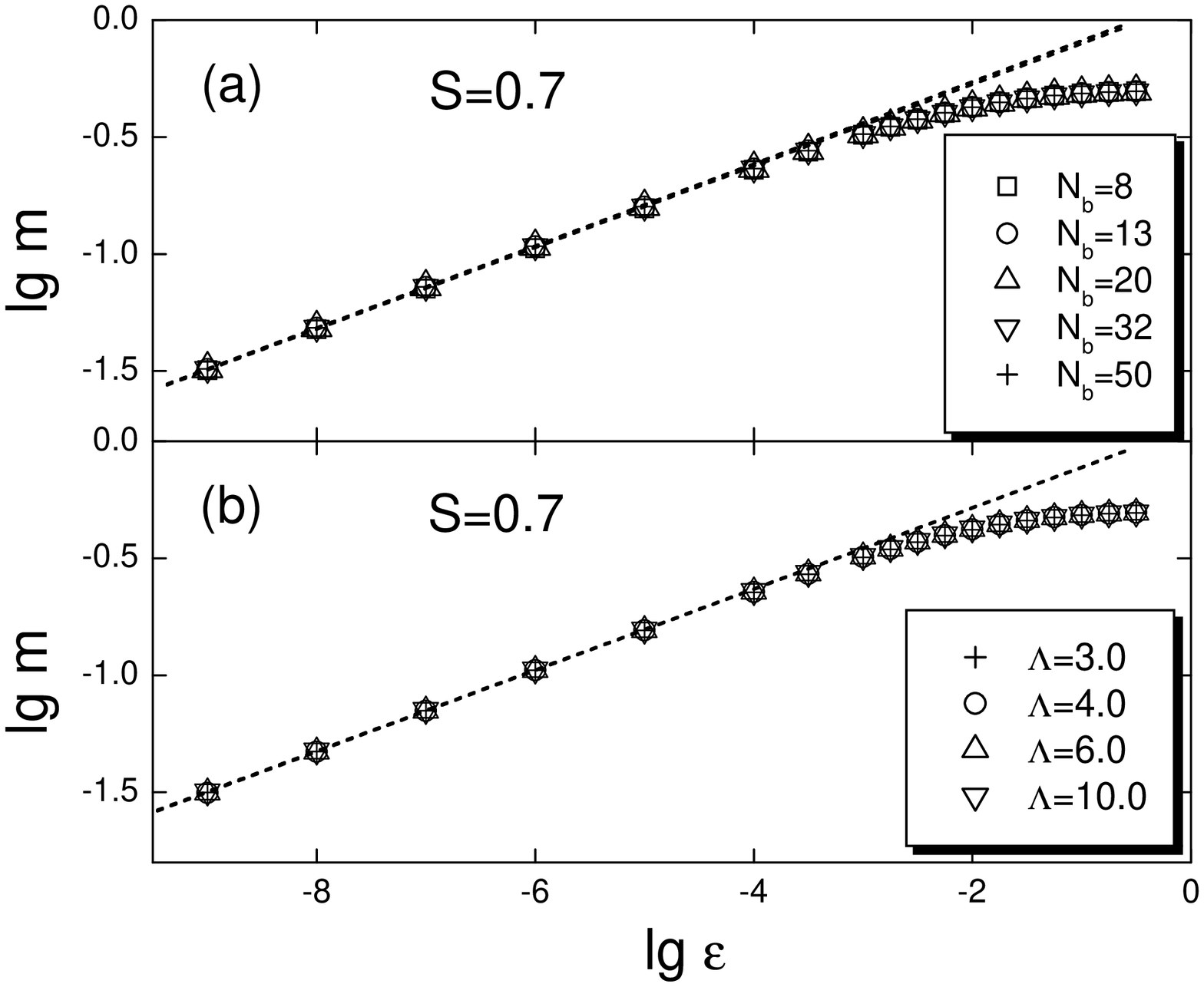}
\vspace*{-1.0cm}
\end{center}
\caption{$\text{lg}m(\tau=0)$ vs. $\text{lg}\epsilon$ at $s=0.7$ for
 (a) different $N_{b}$'s and $\Lambda=10.0$, and
(b) different $\Lambda$'s and $N_{b}=8$. Symbols are BNRG data and
the dashed lines are power law fit, giving the average slopes $0.175$ and $0.173$
for (a) and (b), respectively. Other parameters are $\Delta=0.1$ and
$M_{s}=80$.}
\end{figure}

In Fig.11, we study the $\Delta$ scaling at $s=0.7$ for $m(\tau,
\epsilon=0)$. The main plot and the bottom right inset show that
$m(\tau, \epsilon=0, \Delta) \propto \Delta^{\eta}\tau^{\beta}$. In
the regime $1/2<s<1$, exact expression for $\beta$ as a function of
$s$ is not known. If we assume that $m(\tau, \epsilon, \Delta)$ has
$(\tau/\Delta^{1-s})$ scaling, as observed in the $s=0.3$ case, it
is then easy to obtain
\begin{equation}
   m(\tau, \epsilon=0, \Delta) \propto \Delta^{-(1-s)\beta}
   \tau^{\beta}, \,\,\,\,\,\, (\tau \ll \tau_{sat}).
\end{equation}
For $s=0.7$, the fitted value of
$\eta$ is $-0.096$, which agrees with $-(1-s)\beta=-0.089$ quite well.
 We also checked other $s$ values in the regime $1/2<s<1$
and found good agreement, confirming the validity of
$\tau/\Delta^{1-s}$ scaling in this case. The relation $\alpha_c(1)
\propto \Delta^{1-s}$ is demonstrated in the top left inset of Fig.
11. $\tau_{sat} \propto \Delta^{1-s}$ is the crossover scale
separating the power-law regime from the saturation regime $m \sim
1/2$.

Similar analysis is carried out for $m(\tau=0, \epsilon, \Delta, x,
w)$. In Figs. 12(a) and 12(b), we plot this function for $s=0.7$ and
$\Delta=0.1$, for various $N_{b}$'s ($\Lambda=10$) and $\Lambda$'s
($N_{b}=8$). No $N_{b}$ or $\Lambda$ scaling is observed, similar to
the curves $m(\tau, \epsilon=0, \Delta, x, w)$ shown in Fig.10.

The $\Delta$ scaling carried out in Fig.13 shows that the exponent
$\delta$ fulfills the expression $(1+s)/(1-s)$, even in the
$1/2<s<1$ regime. This is consistent with the previous findings of
the hyperscaling relation $\delta=(1+x)/(1-x)$ and
$x=s$.~\cite{Vojta1} In the inset of Fig.13,
$\text{lg}m-\text{lg}\Delta$ is shown for fixed $\epsilon$,
disclosing a power law behavior consistent with $m(\tau=0, \epsilon,
\Delta) \propto \Delta^{-(1-s)/(1+s)}$. This implies the
$\epsilon/\Delta$ scaling in the regime $1/2<s<1$. Therefore, we
have
\begin{equation}
   m(\tau=0, \epsilon, \Delta, w, w) \propto \left(
   \frac{\epsilon}{\Delta}\right)^{\frac{1-s}{1+s}} \,\,\,\,\, (\epsilon \ll \epsilon_{sat}).
\end{equation}
Here $\epsilon_{sat} \propto \Delta$ is the crossover scale
separating the power law regime from the saturation regime $m \sim
1/2$.

\begin{figure}[t!]
\begin{center}
\includegraphics[width=2.5in, height=3.6in, angle=270]{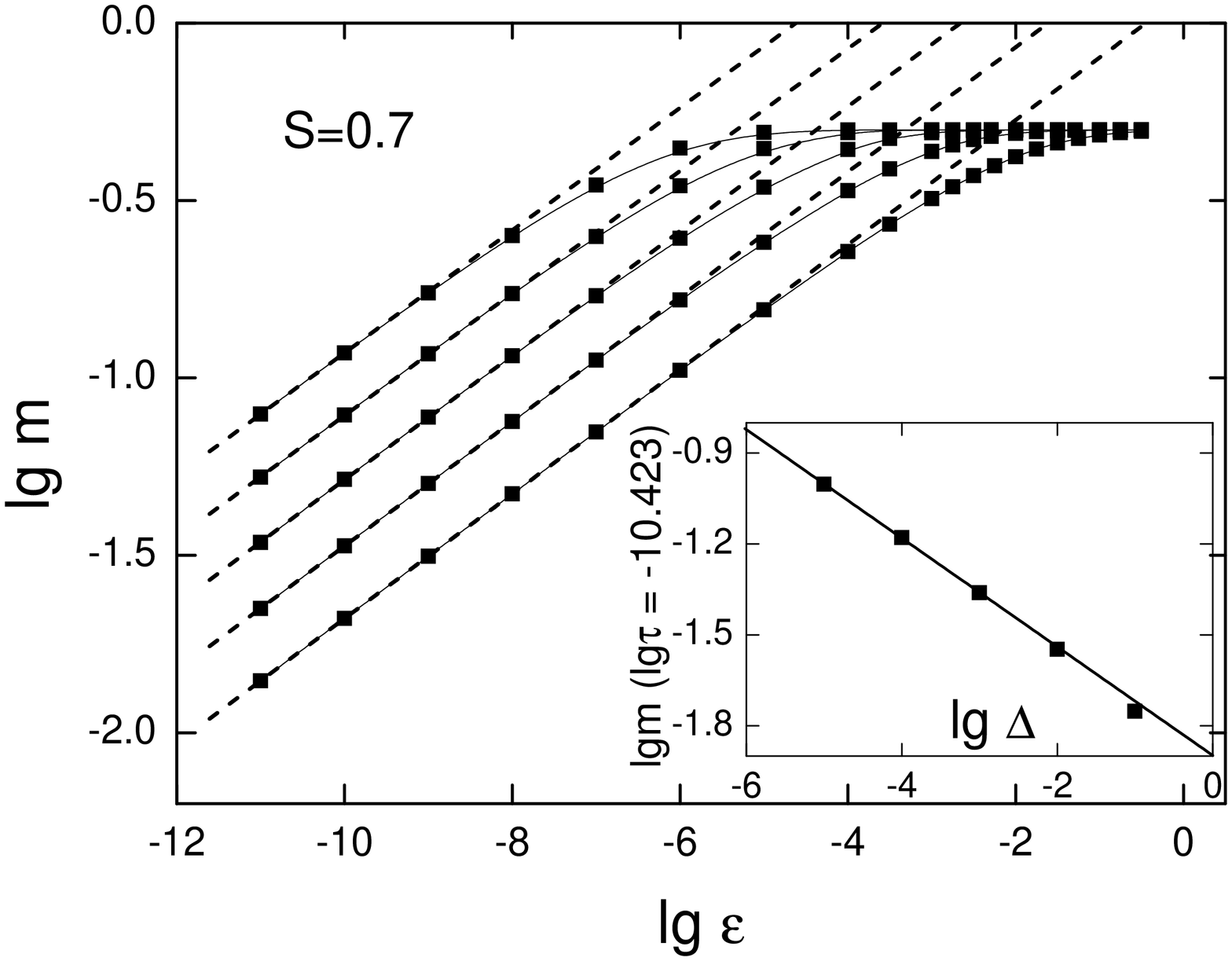}
\vspace*{-1.0cm}
\end{center}
\caption{$\text{lg}m(\tau=0)$ vs. $\text{lg}\epsilon$ at $s=0.7$ for
different $\Delta$'s.
 Symbols are BNRG data and
dashed lines are power law fit. From bottom to top,
$\Delta=10^{-1},10^{-2}, 10^{-3},10^{-4},10^{-5}$, respectively.
Inset:  $\text{lg}m(\tau=0)$ at $\text{lg}\epsilon=-10.423$ as a
function of $\text{lg}\Delta$. The solid line is a power law fit in
small $\Delta$ limit and the slope $-0.179$ is close to
$-(1-s)/(1+s)=-0.177$, as expected from $\epsilon/\Delta$ scaling.
Other parameters are $\Lambda=4.0$, $N_{b}=8$, and $M_{s}=80$.}
\end{figure}

\subsection{$s=0$ and $s=1/2$}

The boundary cases $s=0$ and $s=1/2$ need some discussions. For
$s=0$, the mean-field solution at finite $N_{b}$ gives [Eq.(12)]
$m(\tau, \epsilon=0) \propto \text{exp}(-\Delta/4\omega_c \tau)$ and
the crossover scale $\tau_{cr}$ becomes independent of $wx$. This
means that $\tau_{cr}$ is large and decreases with $wx$ at
subleading order. We carried out the BNRG calculation for
 $s=0$ with $10^{-6} \leqslant \Delta  \leqslant 10^{-2}$ and $10 \leqslant N_{b} \leqslant 50$.
 Using the exponential fit, we obtain $\alpha_c=0$ within an error less than $10^{-7}$.
For a fixed $N_{b}$, $m(\tau, \epsilon=0)$ shows a nice exponential
behavior $m \propto \text{exp}(- c/ \tau)$ with $c \approx (0.27 \pm
0.02)\Delta$. This is consistent with the mean-field expression
(12). For the prefactor, we observe the $x^{-1/2}$ behavior using $8
< N_{b} \leqslant 50$, and find the $-1/2$ exponent of $w$ using
$\Lambda < 2.5$ and $M_{s}> 200$. However, we also find a perfect
$\tau/\Delta$ scaling at $s=0$, being different from Eq. (12). By
combining these results, we expect the crossover scale $\tau_{cr}
\propto -\Delta/\text{ln}(wx)$. As for $m(\tau=0, \epsilon)$, BNRG
gives the exact result $m(\tau=0, \epsilon)=-\epsilon/(2 \Delta)$
since at $\tau=\alpha=0$ the spin is decoupled from the bath.

For $s=1/2$, contrary to $s=0$, the powers of $wx$ in the
expressions for $\tau_{cr}$ and $\epsilon_{cr}$ diverge, leading to
the vanishing of the latter for $wx <1$. Therefore, the classical
regimes in $m(\tau, \epsilon=0)$ and $m(\tau=0, \epsilon)$ expand to
the zero $\tau$ or $\epsilon$ limit, even for finite $N_{b}$'s and
$\Lambda > 1$. Indeed, using $N_{b}=8$ and $\Lambda =4$, BNRG gives
accurate $\beta_{MF}$ and $\delta_{MF}$ at $s=1/2$, as shown in
Fig.14(a) and (b), respectively. For $s=1/2$, corrections to power
laws in $m(\tau, \epsilon=0)$ and $m(\tau=0, \epsilon)$ may arise in
the subleading order in which
 $x$ and $w$ may play some roles. However, such effects are difficult to observe
 in the BNRG data because of numerical errors and fitting errors.

\section{Summary and Discussion}

\begin{figure}[t!]
\begin{center}
\includegraphics[width=3.1in, height=3.7in, angle=270]{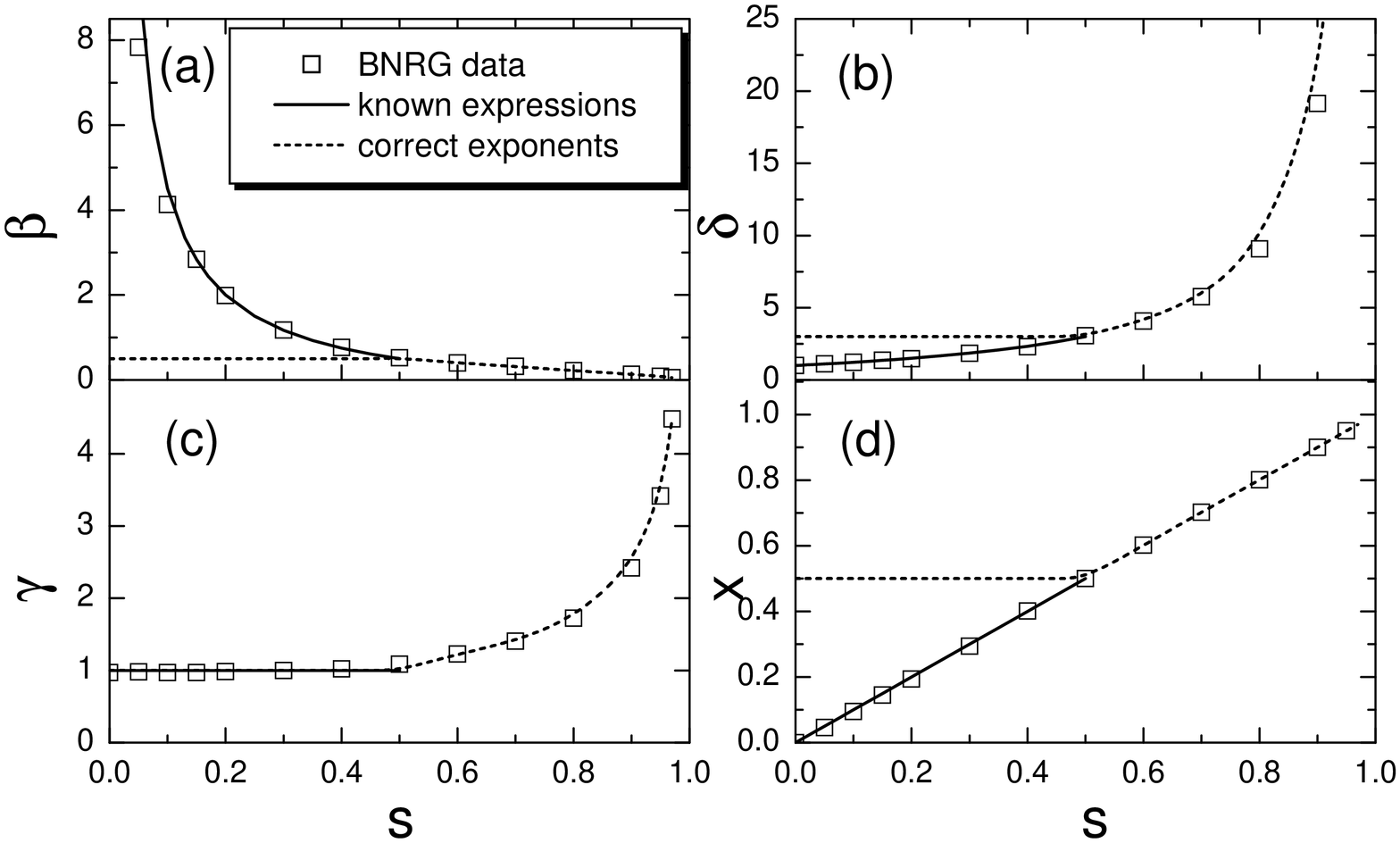}
\vspace*{-1.5cm}
\end{center}
\caption{Critical exponents of the spin-boson model as functions of
$s$: (a) $\beta$, (b) $\delta$, (c) $\gamma$, and (d) $x$. The
squares are the exponents fitted from BNRG data with finite $N_{b}$.
The solid lines are $\beta_{NRG}=(1-s)/(2s)$,
$\delta_{NRG}=(1+s)/(1-s)$, $\gamma_{NRG}=1$, and $x_{NRG}=s$,
respectively. The dashed lines are the correct exponents free from
boson state truncation error and the mass flow error.}
\end{figure}

In Fig.14, we summarize the exponents $\beta$, $\delta$, $\gamma$
and $x$ for the spin-boson model in the whole regime $0<s<1$. The
squares in the figure show the naive BNRG results obtained using
finite $N_{b}$. The solid lines for $\beta$ and $\delta$ in the
regime $0<s<1/2$, as shown in Fig.14(a) and (b), are
$\beta=2s/(1-s)$ and $\delta=(1+s)/(1-s)$, the exact solution of the
mean-field Hamiltonian with finite $N_{b}$. They agree well with the
BNRG data (squares). The dashed lines are the correct values
extracted from the scaling analysis. They deviate significantly from
the BNRG results in $0 \leqslant s < 1/2$. In the regime $1/2
\leqslant s <1$, the boson state truncation error and mass flow
error still exist but they do not influence the exponents.

We also studied the magnetic susceptibility $\chi$ using BNRG in the
regime $0 \leqslant s \leqslant 1/2$ (squares in Fig.14(c)). It is
found that $\chi(\tau, T=0)$ curve is independent of $w$ and $x$ and
$\gamma=1$ holds at high precision. For the $\Delta$ dependence,
BNRG calculations for various $s$ in the regime
 $0 \leqslant s <1$ confirm the scaling form $\chi \propto \Delta^{-1}
(\tau/\Delta^{1-s})^{-1}$, being consistent with $m(\tau, \epsilon,
\Delta)=f(\tau/\Delta^{1-s}, \epsilon/\Delta)$. In the mean-field
theory, the factors containing $wx$ cancel each other and one
obtains $\chi \propto \Delta^{-1} (\tau/\alpha_{c}(1))^{-1}$,
independent of $wx$. It covers the BNRG result when $\alpha_c(1)$ is
replaced by $\Delta^{1-s}$. In the regime $1/2<s<1$, the BNRG result
for $\gamma$ follows a nonclassical curve [squares and the dashed
line in Fig. 14(c)], the exact expression of which is not known yet.
It is related to $\beta$ and $\delta$ through hyperscaling
relations.

The temperature dependence $\chi(T)$ at $\tau=0$ is governed by the exponent $x$. We
did not find any observable shift in $\chi(\tau=0, T)$ with the
changing of $N_{b}$ or $\Lambda$ in the range $8< N_{b} <20$ and
$2< \Lambda < 10$. We always obtain $x=s$ in the regime $0<s<1/2$.
 This indicates that the mass-flow problem is deeply rooted in the NRG algorithm and can not
be solved by the scaling method that we discuss in this paper.
According to the analysis of the mass-flow problem in NRG,
~\cite{Vojta2,Vojta3} $x=1/2$ should hold in the regime $0 \leqslant
s \leqslant 1/2$ and $x=s$ in $1/2<s<1$.
 This conclusion together with the BNRG result are summarized in Fig. 14(d) for completeness.

To conclude, we confirm the statements in Ref.~\onlinecite{Vojta2}
that for the spin-boson model, the critical exponents $\beta$ and
$\delta$ are classical in the regime $0 \leqslant s \leqslant 1/2$,
while are nonclassical in $1/2<s<1$. BNRG with finite $N_{b}$
produces artificial interacting critical fixed point in the regime
$0 \leqslant s <1/2$, and the extracted $\beta$ and $\delta$ are
incorrect. In the regime
 $1/2 \leqslant s <1$, boson-state truncation does not play a role and BNRG results are
 correct. Scaling analysis with respect to $x=1/N_{b}$ and $w=\Lambda-1$
 can be used as a supplement to ordinary BNRG to overcome the boson-state
  truncation error. It is also observed that the order parameter has the scaling form
$m(\tau, \epsilon, \Delta)= f(\tau/\Delta^{1-s}, \epsilon/\Delta)$
in the whole regime $0 \leqslant s<1$.

Two features of the BNRG results can be understood in terms of the
mean-field Hamiltonian. First, as shown in Eqs. (31) and (33), the
boson-state truncation error disappears in the limit $\Lambda=1$.
Second, the truncation $N_{b}$ influences only the exponents but not
the critical coupling $\alpha_c$. In the mean-field Hamiltonian
[Eqs. (6)-(8)], the $n$th boson mode has a displacement $\gamma_n /
\xi_n$ and its coupling strength to the spin is proportional to
$\gamma_n$. Both $\gamma_n$ and $\xi_n$ decay exponentially with
increasing $n$, but the ratio  $\gamma_n / \xi_n$ diverges as
$\Lambda^{(1-s)n/2}$. The low-energy modes (with large $n$) are more
displaced and need more states to describe. Therefore, truncation of
the state influences mostly the low-energy modes and not the
high-energy modes. In the limit $\Lambda=1$, however, there is no
divergence in the displacement and all the modes can be described
equally well by a finite $N_{b}$, hence the truncation error
disappears. Since the critical exponents are closely related to the
energy distribution and the displacement behavior of the low-energy
boson modes, they are susceptible to the truncation. While the
critical coupling is dominated by the integral properties of the
spectrum in which the high-energy modes have more weights, it is
thus only very weakly influenced by the truncation.

Our results support the validity of quantum-to-classical mapping for
the spin-boson model in the regime $0 \leqslant s < 1/2$. Note that
our picture of $m(\tau)$ for the truncated spin-boson model is
consistent with the results of exact diagonalization study in
Ref.~\onlinecite{Alvermann1}. There, although a different basis set
is used, in principle, the boson-state truncation error also exists
and the nonclassical exponents should be present in the low-energy
limit. However, the fitting of data relatively far away from the
critical point (note the linear scale in Fig. 3 of
Ref.~\onlinecite{Alvermann1}) coincidentally misses the nonclassical
regime and produces the correct classical exponent $\beta_{MF}$.

For other quantum impurity models such as the Bose-Fermi-Kondo model
(BFKM) model~\cite{Si1, Zhu1, Zhu2} and its anisotropic version, the
Ising-BFKM,~\cite{Glossop1} similar quantum-to-classical mapping
arguments supports Gaussian fixed points in the regime $0 \leqslant
s <1/2$. Correspondingly, one expects $\beta=1/2$, $\delta=3$ and
$x=1/2$. Neither hyperscaling relations nor $\omega/T$ scaling
should hold. However, variouls studies using BNRG~\cite{Glossop1},
QMC simulation~\cite{Kirchner2, Kirchner3,Kirchner4},
$\epsilon$-expansion~\cite{Zhu1, Si2} and the large-N limit
analysis~\cite{Si1, Zhu2, Burdin1} all point to the failure of the
quantum-to-classical mapping in these systems: they obtained
interacting fixed point and nonclassical exponents. Especially, for
the anisotropic BFKM with a sub-Ohmic boson bath, the exponents are
found to be the same as $\beta_{NRG}$ and $\delta_{NRG}$ of the
spin-boson model.

In light of the present study, it is possible that the same boson
state truncation error may exist in the BNRG study of the Ising BFKM
in Ref.~\onlinecite{Glossop1} and lead to incorrect $\beta$ and
$\delta$ in the regime $0<s<1/2$.
 Note that the boson state truncation problem
cannot be remedied by going to smaller energy
scales.~\cite{Glossop1,Kirchner4} Since the fermion bath in the
Ising BFKM can be integrated out to produce an additional boson bath
with Ohmic spectrum, the critical behavior should be dominated by
the sub-Ohmic one, leading to the expectation that it belongs to the
same universality class as the spin-boson model. In the quantum
Monte Carlo simulations, it was observed that the truncation of
correlations in the imaginary time axis is the key to produce the
interacting critical point.~\cite{Kirchner4} Otherwise Gaussian
behavior will obtain. These may be signatures that the
quantum-to-classical mapping holds also for the Ising BFKM in $0
\leqslant s < 1/2$. It is straightforward to examine this statement
using BNRG supplemented with the $N_{b}$ scaling.

As far as the isotropic BFKM is concerned, the situation seems
different. Here, the symmetry is different from the spin-boson model
and the Berry phase effect is claimed to be
nontrivial.~\cite{Kirchner5} Much work has been done for this model,
supporting the failure of the quantum-to-classical mapping in the
sub-Ohmic regime.~\cite{Zhu2, Kirchner3, Kirchner5} Due to the very
subtle nature of this problem, however, exact numerical studies are
still desirable as a further confirmation. Both the isotropic and
the anisotropic BFKM are at the core of studying the Kondo lattice
with Ruderman-Kittel-Kasuya-Yoshida (RKKY) interactions using the
extended dynamical mean-field
theory.~\cite{Si1,Si2,Glossop2,Zhu3,Zarand1,Zhu4,Grempel1}
Therefore, any definite conclusions concerning these impurity models
will have important impact on the understanding of the competition
between Kondo screening and the antiferromagnetic state in the
heavy-fermions metals.~\cite{Si1, Gegenwart1}

In summary, we carried out systematic BNRG studies for the
spin-boson model in the sub-Ohmic regime, supplemented with the
scaling analysis for the boson state truncation $x=1/N_{b}$ and the
logarithmic discretization parameter $w=\Lambda-1$. For $0<s<1/2$,
the function $m(\tau, \epsilon=0, \Delta, x, w)$ [$m(\tau=0,
\epsilon, \Delta, x, w)$] is shown to bear a multiple power form in
the small-$\tau$ (-$\epsilon$) limit. Classical exponent $\beta=1/2$
($\delta=3$) is identified in the regime $\tau \gg \tau_{cr}$
($\epsilon \gg \epsilon_{cr}$), agreeing with the conclusion from
the quantum-to-classical mapping. The crossover scale $\tau_{cr}$
($\epsilon_{cr}$) goes to zero in the small $x$ or $w$ limits in a
power law. This presents a scenario of how the boson-state
truncation error disappears in the limit $N_{b} \rightarrow \infty$.
The observation that $x$ and $w$ always appear as a product $(wx)$
indicates that, in the regime $0 \leqslant s < 1/2$, the boson-state
truncation invalidates the logarithmic discretization scheme, which
is the basis of NRG. Independent of the issue of boson-state
truncation, we also find that the scaling form for the order
parameter $m(\tau, \epsilon, \Delta) =f(\tau/\Delta^{1-s},
\epsilon/\Delta)$ in the whole regime $0 \leqslant s<1$.

\section{Acknowledgments}
The authors acknowledge helpful discussions with Ralf Bulla,
Hsiu-Hau Lin, and Matthias Vojta. This work is supported by the 973
Program of China under Grant No. 2007CB925004 and by National
Natural Science Foundation of China under Grant No.11074302.

\begin{appendix}
\section{Critical Exponents of the Mean-Field Spin-Boson Model}

In this appendix, we present the calculation of the critical
exponents $\beta$, $\delta$, and $\gamma$ for the mean-field
spin-boson Hamiltonian.

The Hamiltonian of the spin-boson model reads as
\begin{equation}
   H_{sb}=-\frac{\Delta}{2} \sigma_x + \frac{\epsilon}{2} \sigma_z
  +\frac{1}{2}\sigma_z  \sum_{i} \lambda_{i}\left(a_{i}+a_{i}^{\dagger} \right)
  + \sum_{i} \omega_{i} a_{i}^{\dagger}a_{i}.
\end{equation}
The mean-field Hamiltonian is (neglecting a constant)
\begin{eqnarray}
  H_{mf}= &&  -\frac{\Delta}{2}\sigma_{x}+ \left[ \frac{\epsilon}{2}+ \frac{1}{2} \sum_{i}\lambda_{i}
   \langle a_{i}+a_{i}^{\dag} \rangle \right] \sigma_{z} \nonumber   \\
    && + \sum_{i}\omega_{i} a_{i}^{\dag}a_{i} + \frac{\langle \sigma_z\rangle}{2}
   \sum_{i}\lambda_{i}\left(a_{i}+a_{i}^{\dag} \right).
\end{eqnarray}
For a finite $N_{b}$, the critical behavior depends severely on the
parameterization scheme for $\{\omega_{i}, \lambda_{i} \}$ as well
as on the discretization scheme. In order to compare the exponents
with the NRG ones, we carry out the logarithmic discretization as in
NRG and get
\begin{eqnarray}
  H_{mf}^{LD}= &&  -\frac{\Delta}{2}\sigma_{x}+ \left[ \frac{\epsilon}{2}+ \frac{1}{2\sqrt{\pi}} \sum_{n=0}^{\infty}\gamma_{n}
   \langle a_{n}+a_{n}^{\dag} \rangle \right] \sigma_{z} \nonumber   \\
    && + \sum_{n=0}^{\infty}\xi_{n}a_{n}^{\dag}a_{n} + \frac{\langle \sigma_z\rangle}{2\sqrt{\pi}}
   \sum_{n=0}^{\infty}\gamma_{n}\left(a_{n}^{\dagger}+a_{n} \right),
\end{eqnarray}
where the superscript LD denotes the logarithmic discretization.
$\gamma_{n}$ and $\xi_{n}$ are obtained from standard
procedure~\cite{Bulla1, Bulla3} and given in Eq.(7) and (8) in the main text. In
Eq.(A3), the spin and boson degrees of freedom decouple and we get
$H_{mf}^{LD}=H_{spin}+H_{boson}$, where
\begin{equation}
   H_{spin}=-\frac{\Delta}{2}\sigma_{x} +
   \frac{1}{2}(\epsilon+d)\sigma_{z},
\end{equation}
and
\begin{equation}
   H_{boson}=\sum_{n=0}^{\infty}\xi_{n}a_{n}^{\dag}a_{n} + \frac{1}{2} \langle \sigma_z\rangle
   \sum_{n=0}^{\infty} \frac{\gamma_{n}}{\sqrt{\pi}} \left(a_{n}^{\dagger}+a_{n}
   \right).
\end{equation}
Here
\begin{equation}
 d=\sum_{n=0}^{\infty} \frac{\gamma_{n}}{\sqrt{\pi}} \langle a_{n}^{\dagger}+ a_{n}
 \rangle.
\end{equation}

For $N_{b}=\infty$, $H_{boson}$ can be solved exactly. For finite
$N_{b}$, it is no longer exactly solvable. In Ref.~\onlinecite{Hou1}, $H_{mf}^{LD}$
is solved numerically and the critical behavior is studied. Here, we
start from the single mode Hamiltonian
\begin{equation}
   H=a^{\dagger}a + \theta(a^{\dagger} + a).
\end{equation}
In the small $\theta$ limit, the ground state is not influenced by
the truncation and we have $\langle a^{\dagger} + a \rangle_{G}=-2
\theta$ and $\langle a^{\dagger} a \rangle_{G}=\theta^2$. In the
large $\theta$ limit, the number of bosons in the ground state is
confined by the boson state truncation $N_{b}$. Therefore in this
limit $\langle a^{\dagger}a \rangle_{G}=N_{b}$. For the displaced
harmonic oscillator, to leading order of $N_{b}$ we have $\langle a
+ a^{\dagger} \rangle_{G}=-2 \sqrt{ \langle a^{\dagger}a
\rangle_{G}}=-2 \sqrt{ N_{b} }$. The crossover scale is determined
by equating the two situations. In summary, we have
\begin{eqnarray}
\langle a^{\dagger}+a \rangle_{G} = \left\{
\begin{array}{lll}
 -2\theta ,\,\,\,\,\,   & (\theta \ll \sqrt{N_{b}} ); \\
&\\
 -2\sqrt{N_{b}} ,\,\,\,\,\, & (\theta \gg \sqrt{N_{b}} ).
\end{array} \right.
\end{eqnarray}
We confirmed this result numerically. Here $\langle ... \rangle_{G}$
denotes the ground state average. $\langle a_{n}^{\dagger}+a_{n}
\rangle_{G}$ has different form in the regimes $n \ll n_{0}$ and $n
\gg n_{0}$. Here $n_{0}$ separates the freely biased mode (small
$n$) from the saturated biased mode (large $n$). We can then solve
$H_{boson}$ approximately and obtain one of the mean-field equations
from Eq.(A6),
\begin{equation}
   d \approx -\frac{1}{\pi}\langle \sigma_z \rangle
   \sum_{n=0}^{n_{0}} \frac{ \gamma_{n}^{2} }{ \xi_{n} } -2
   \sqrt{\frac{N_{b}}{\pi}} \sum_{n=n_{0}+1}^{\infty} \gamma_{n},
\end{equation}
with $n_{0}$ determined by equating $\sqrt{N_{b}}$ with the
effective bias,
\begin{equation}
   \frac{1}{2} \langle \sigma_z \rangle
   \frac{\gamma_{n_{0}}}{\sqrt{\pi}\xi_{n_{0}}} =\sqrt{N_{b}}.
\end{equation}
Introducing $x=1/N_{b}$ and $m=\langle \sigma_z \rangle/2$ and
summing over $n$ in Eq.(A9) produces
\begin{equation}
 d=a_{1}(\Lambda) \alpha m + \left[ a_{2}(\Lambda) + b_{1}(\Lambda) \right] \alpha^{\frac{1}{1-s}}
 x^{\frac{s}{1-s}} m^{\frac{1+s}{1-s}}.
\end{equation}
The parameters $a_{1}(\Lambda)$, $a_{2}(\Lambda)$, and
$b_{1}(\Lambda)$ read
\begin{equation}
   a_{1}(\Lambda)=-\frac{4(2+s)}{(1+s)^{2}}\frac{\left[ 1-\Lambda^{-(1+s)} \right]^{2} }{ \left[1-\Lambda^{-s} \right]\left[1-\Lambda^{-(2+s)}
   \right]} \omega_c,
\end{equation}
\begin{eqnarray}
  a_{2}(\Lambda)&=&2^{\frac{2-s}{1-s}}(1+s)^{-\frac{2+s}{1-s}}(2+s)^{\frac{1+s}{1-s}}\left[1-\Lambda^{-(2+s)}\right]^{-\frac{1+s}{1-s}} \nonumber \\
&&
   \left[1-\Lambda^{-(1+s)}\right]^{\frac{2+s}{1-s}}\frac{1}{\Lambda^{s}-1} \omega_c,
\end{eqnarray}
and
\begin{equation}
  b_{1}(\Lambda)=-a_{2}(\Lambda)\frac{\Lambda^{s}-1}{\Lambda^{(1+s)/2}-1}.
\end{equation}

The other mean-field equation from solving $H_{spin}$ is
\begin{equation}
    m=\frac{1}{2}\frac{\Delta^2-\left[ \epsilon+d+\sqrt{\Delta^2+(\epsilon+d)^2}\right]^2}{\Delta^2+\left[\epsilon+d+\sqrt{\Delta^2+(\epsilon+d)^{2}}
    \right]^2}.
\end{equation}
Near the critical point, it reduces to
\begin{equation}
   m=-\frac{\epsilon+d}{2\Delta}+ \frac{1}{4}\left( \frac{\epsilon+d}{\Delta}
   \right)^{3} + O\left( \frac{\epsilon+d}{\Delta}
   \right)^{4}.
\end{equation}
Putting Eq.A(11) into the above equation and keeping only the
leading term of each type, one gets
\begin{eqnarray}
m= && -\frac{\epsilon}{2\Delta} -\frac{ a_{1}(\Lambda) \alpha
m}{2 \Delta} + \frac{1}{4 \Delta^{3}} \left[a_{1}(\Lambda) \alpha m
 \right]^{3}   \nonumber \\
&&
-\frac{a_{2}(\Lambda)+b_{1}(\Lambda)}{2\Delta}\alpha^{\frac{1}{1-s}}x^{\frac{s}{1-s}}
m^{\frac{1+s}{1-s}}.
\end{eqnarray}
From this equation the critical coupling strength is obtained as
$\alpha_c(\Lambda)=-2 \Delta/a_{1}(\Lambda)$. It is remarkable that it
is independent of the boson state truncation $N_{b}$. Introducing
$\tau=\alpha- \alpha_c(\Lambda)$ to measure the distance to the
critical point, we get the self-consistent equation as
\begin{eqnarray}
  &&-\frac{\epsilon}{\Delta} -\frac{a_{2}(\Lambda)+b_{1}(\Lambda)}{\Delta}
\left[\alpha_c(\Lambda)\right]^{\frac{1}{1-s}} x^{\frac{s}{1-s}}
m^{\frac{1+s}{1-s}}    \nonumber \\
&& + \frac{2 \tau}{\alpha_c(\Lambda)}m-4 m^{3}=0.
\end{eqnarray}
To investigate the role of $\Lambda$, we define $w=\Lambda-1$ and
expand $\alpha_{c}(\Lambda)$, $a_{2}(\Lambda)$ and $b_{1}(\Lambda)$
to leading order of $w$. Finally we get the self-consistent equation
in terms of $\tau$, $\epsilon$, $\Delta$, $w$, and $x$ as (keeping
$\alpha_c(\Lambda)$ in the definition of $\tau$ unchanged)
\begin{equation}
 \frac{2 \tau}{\alpha_c(1)} m - 4 m^{3}
 -\frac{\epsilon}{\Delta}
 - c\frac{\omega_{c}}{\Delta} \left[\alpha_c(1)
 \right]^{\frac{1}{1-s}} (wx)^{\frac{s}{1-s}} m^{\frac{1+s}{1-s}}=0.
\end{equation}
Here $\alpha_{c}(1)=\Delta s/(2 \omega_c)$ is the critical $\alpha$ value for $\Lambda=1$.
$c=2^{(2-s)/(1-s)}\left[1/s - 2/(1+s) \right]$ is a constant. It is
noted here that the anomalous term involving the $(wx)^{s/(1-s)}$
comes solely from the factor $a_{2}(\Lambda)+b_{1}(\Lambda)$,
instead of from expansions of $\alpha_c(\Lambda)$ in Eq.(A18).
Solving this equation in various limits produces the analytical
expressions Eqs.(9)-(16).

Eq.(A19) reduces to standard mean-field equation and gives
$\beta=1/2$, $\delta=3$, when the $m^{3}$ term dominates over the
anomalous term $(wx)^{s/(1-s)}m^{(1+s)/(1-s)}$. This is the case
when $m$ is small in the regime $1/2<s<1$, or when $wx \ll m$ in the
regime $0<s<1/2$. The anomalous term will dominate over the $m^{3}$
term and give $\beta=(1-s)/(2s)$ and $\delta=(1+s)/(1-s)$, when $m
\rightarrow 0$ in the regime $0<s<1/2$. Considering different
regimes of $\tau$ and $\epsilon$, we get the expressions (9) and
(10) and (13) and (14). The crossover scales $\tau_{cr}$ [Eq. (11)]
and $\epsilon_{cr}$ [Eq.(15)] are obtained by equating the
expressions in the classical and nonclassical regimes.

At $s=0$, Eq. A(10) is replaced with
\begin{eqnarray}
n_{0}=\frac{1}{\text{ln}\Lambda}\text{ln}\left[\frac{(1+\Lambda^{-1})^2}{1-\Lambda^{-1}}
\frac{1}{8 \alpha x m^{2} } \right].
\end{eqnarray}
The summation in Eq.A(9) produces
\begin{eqnarray}
d=&& -8\alpha m \omega_c
\frac{1-\Lambda^{-1}}{(1+\Lambda^{-1}) \text{ln}\Lambda}\text{ln}
\left[ \frac{\Lambda (1+\Lambda^{-1})^{2} }{8(1-\Lambda^{-1}) x
\alpha m^{2} } \right] \nonumber \\
&& -8\alpha m \omega_c
\frac{\Lambda^{-1/2}+\Lambda^{-1}}{1+\Lambda^{-1}}.
\end{eqnarray}
This gives $\alpha_{c}=0$. Expanding $d$ at $\Lambda=1$ to
leading order and combining Eq.(A16), we get the order
parameter for $s=0$,
\begin{equation}
   m(\tau, \epsilon=0, \Delta, x, w)=c \frac{1}{2}(w x \tau)^{-\frac{1}{2}} e^{-\frac{\Delta}{4\omega_{c}
   \tau}},
\end{equation}
and
\begin{equation}
   m(\tau=0, \epsilon, \Delta, x, w)= - \frac{\epsilon}{2\Delta}.
\end{equation}
Here, $c$ is a constant independent of $\Delta$, $x$, and $w$. Note
that if one takes the $x \rightarrow 0$ limit first and then takes
the limit $s \rightarrow 0$, Eq.(A22) becomes
\begin{eqnarray}
m(\tau, \epsilon=0, \Delta, x=0, w) = \left\{
\begin{array}{lll}
 1/2 ,\,\,\,\,\,   & (\tau > 0); \\
&\\
 0 ,\,\,\,\,\,  & (\tau =0).
\end{array} \right.
\end{eqnarray}

The magnetic susceptibility $\chi$ can be obtained by taking
derivative on both sides of Eq. (A19) with respect to $\epsilon$.
This leads to the expression
\begin{eqnarray}
 && - \Delta \chi =    \nonumber \\
 &&  \left[\frac{c(1+s)}{1-s}\frac{\omega_c}{\Delta} \left[\alpha_{c}(1) \right]
 ^{\frac{1}{1-s}}(wx)^{\frac{s}{1-s}}m^{\frac{2s}{1-s}} +12 m^{2} - \frac{2\tau}{\alpha_c(1)} \right]^{-1}. \nonumber \\
 &&
\end{eqnarray}
Here, $c=2^{(2-s)/(1-s)}\left[1/s - 2/(1+s) \right]$. Analyzing this
equation in different $\tau$ regimes, we obtain $\chi(\tau, \Delta)
= -\alpha_c(1)/(4 \Delta \tau)$ in the regime $1/2 \leqslant s < 1$,
and
\begin{eqnarray}
\chi(\tau, \Delta) = \left\{
\begin{array}{lll}
 -\frac{\alpha_{c}(1)}{4\Delta \tau}      \,\,   & (\tau \gg \tau_{cr}),  \\
&\\
 -\frac{1-s}{4s}\frac{\alpha_c(1)}{\Delta \tau}   \,\, & (\tau \ll \tau_{cr}).
\end{array} \right.
\end{eqnarray}
in the regime $0 \leqslant s < 1/2$. This gives the exponent
$\gamma=1$, independent of $N_{b}$ and $w$.

We also studied the critical behavior of $m(\tau, \epsilon, \Delta,
x)$ using other parametrization and discretization schemes for the
bath spectrum. It is found that for finite $N_{b}$, the critical
exponents $\beta$ and $\delta$ are strongly dependent on the scheme.
For $N_{b}=\infty$, they all reduce to the mean-field values.

\end{appendix}

\end{document}